\documentclass[preprint2]{aastex}

\newcommand{\neonteff}{[\ion{Ne}{3}]\,15.6\,$\rm \mu m$/[\ion{Ne}{2}]\,12.8\,$\rm \mu m$}
\newcommand{\snrate}{\ensuremath{\nu_{\rm SN}}}
\newcommand{\lir}{\ensuremath{L_{\rm IR}}}
\newcommand{\lbol}{\ensuremath{L_{\rm bol}}}
\newcommand{\llyc}{\ensuremath{L_{Lyc}}}
\newcommand{\etal}{{\em et al.}\, }
\newcommand{\slyc}{\ensuremath{_{Lyc}}}
\newcommand{\mup}{\ensuremath{m_{\rm up}}}
\newcommand{\mlow}{\ensuremath{m_{\rm low}}}
\newcommand{\tburst}{\ensuremath{t_{\rm b}}}
\newcommand{\tsc}{\ensuremath{t_{\rm sc}}}
\newcommand{\um}{\ensuremath{\mu}m}

\newcommand{\msolar}{\ensuremath{M_\odot}}
\newcommand{\zsolar}{\ensuremath{Z_\odot}}
\newcommand{\lsolar}{\ensuremath{L_\odot}}

\slugcomment{Accepted for publication in the Astrophysical Journal}
\shortauthors{Thornley \etal}
\shorttitle{Mid-infrared line emission from starburst galaxies}

\begin{document}

\title{Massive star formation and evolution in starburst galaxies:
  mid-infrared spectroscopy with ISO-SWS
  \footnote{Based on observations with ISO, an ESA project with
  instruments funded by ESA Member States (especially the PI
  countries: France, Germany, the Netherlands, and the United Kingdom)
  and with the participation of ISAS and NASA. The SWS is a joint
  project of SRON and MPE.}}

\author{Michele D. Thornley\altaffilmark{1}, Natascha M. F\"orster
Schreiber\altaffilmark{2}, Dieter Lutz, Reinhard Genzel, Henrik
W. W. Spoon\altaffilmark{3}, Dietmar Kunze}
\affil{Max-Planck-Institut f\"ur extraterrestrische Physik (MPE),
Postfach 1603, D-85740 Garching bei M\"unchen, Germany}
\and
\author{Amiel Sternberg}
\affil{School of Physics and Astronomy, Tel Aviv University, Ramat
Aviv, Tel Aviv 69978, Israel}

\altaffiltext{1}{Current address: National
Radio Astronomy Observatory, 520 Edgemont Rd., Charlottesville, VA
22903 USA; mthornle@nrao.edu} 
\altaffiltext{2}{Current address: CEA/DAPNIA/SAp, Centre
d'Etudes de Saclay, Orme-des-Merisiers B\^at. 709, F-91191
Gif-sur-Yvette, France} 
\altaffiltext{3}{Current address: European
Southern Observatory, Karl-Schwarzschild-Strasse 2, D-85748 Garching
bei M\"unchen, Germany}

\begin{abstract}
  We present new ISO-SWS data for a sample of 27 starburst galaxies,
  and with these data examine the issues of formation and evolution of
  the most massive stars in starburst galaxies.  Using starburst
  models which incorporate time evolution, new stellar atmosphere
  models for massive stars, and a starburst model
  geometry derived from observations of the prototypical starburst
  M82, we model the integrated mid-infrared line ratio
  [Ne~III](15.6\um)/[Ne~II](12.8\um). This line ratio is sensitive to
  the hardness of the stellar energy distribution and therefore to the
  most massive stars present.

  We conclude from our models, with consideration of recent
determinations of the stellar census in local, high-mass star forming
regions, that the [Ne~III]/[Ne~II] ratios we measure are consistent
with the formation of massive ($\sim$ 50-100\msolar) stars in most
starbursts.  In this framework, the low nebular excitation inferred
from the measured line ratios can be attributed to aging effects.  By
including estimates of the ratio of infrared-to-Lyman continuum
luminosity for the galaxies in our sample, we further find that most
starbursts are relatively short-lived (10$^6$-10$^7$ years), only a
few O-star lifetimes.  We discuss a possible cause of such short
events: the effectiveness of stellar winds and supernovae in
destroying the starburst environment.

\end{abstract}
\keywords{galaxies:starburst---infrared:galaxies---techniques:spectroscopic---stars:formation,
evolution, and atmospheres}

\section{Introduction} \label{Sintro}

Starburst galaxies constitute an important class of extragalactic
objects.  They contribute a significant fraction of the total
high-mass star formation in the local universe ({\em e.g.} Soifer
\etal 1987; Gallego \etal 1995), and about 25\% of the high-mass star
formation within 10 Mpc occurs in four starburst galaxies (Heckman
1998).  At intermediate and high redshifts, important populations of
galaxies are observed which exhibit properties indicating intense star
formation activity ({\em e.g.} Colless \etal 1994; Babul \& Ferguson
1996; Steidel \etal 1996; Lowenthal \etal 1997).  A strong correlation
between the most prodigious starbursts and interaction or merger
events ({\em e.g.}  Condon \etal 1982; Kennicutt \etal 1987; Telesco,
Wolstencroft \& Done 1988; Joseph 1990) further emphasizes the
importance of starbursts in galaxy evolution.

Despite the obvious importance of starbursts, a canvas of recent
starburst galaxy studies suggests that a full ``prescription'' for
these events has yet to be written.  Discrepancies exist even for the
most massive stars formed, which represent the largest energy
contributions. For instance, it has been suggested by some authors
that the production of the highest mass stars may be suppressed in
starburst galaxies, based on low measured values of diagnostic line
ratios such as He I (2.06\um)/ HI(Br$\gamma$) and [Ne II]/[Ar III]
detected in well-known starburst sources.  These results have been
interpreted as possibly indicating severe upper mass cutoffs to the
initial mass function (IMF), some as low as $\sim$25-30\msolar~({\em
e.g.}  Puxley \etal 1989; Doyon \etal 1994; Doherty \etal 1995;
Achtermann \& Lacy 1995; Beck, Kelly, \& Lacy 1997). 

A lack of massive stars in starbursts is difficult to reconcile with a
growing body of evidence for the formation of very massive stars in
nearby regions of active star-formation.  In recent studies of local
high-mass star-forming regions at both high and low metallicity, stars
up to at least 100~\msolar~are observed; these regions include the
Galactic Center (e.g. Krabbe \etal 1995; Najarro \etal 1997; Serabyn,
Shupe, \& Figer 1998; Figer \etal 1998), the Galactic star-forming
region NGC\,3603 (Drissen \etal 1995; Eisenhauer \etal 1998), and the
R136 cluster at the center of 30 Doradus (e.g. Hunter \etal 1995,
Massey \& Hunter 1998).  There is also a growing body of indirect
evidence for the presence of very massive stars in starbursts.  {\em
HST} optical imaging has revealed the presence of young compact
``super star clusters'' in several nearby starburst galaxies,
including M\,82 (O'Connell \etal 1995), NGC\,4038/4039 (Whitmore \&
Schweizer 1995), NGC 5253 (Meurer \etal 1995), He 2-10 (Conti \& Vacca
1994), NGC\,1569 and NGC\,1705 (O'Connell, Gallagher, \& Hunter 1994; Ho \& Filippenko
1996; Sternberg 1998), and NGC\,1140 (Hunter, O'Connell \& Gallagher
1994).  These super star clusters have extreme luminosities, and in
some cases bright emission features characteristic of Wolf-Rayet stars
(e.g., He II 4696\AA~emission); such properties are difficult to
explain without a large contribution from very massive stars
(M$\ge$60\msolar; see e.g., Gonzalez Delgado \etal 1997; Schaerer
\etal 1997).

Several authors have recently argued that observed low-excitation
nebular line ratios in starbursts may be due to aging effects rather
than a severe upper mass cutoff ({\em e.g.}  Rieke \etal 1993; Genzel,
Hollenbach, \& Townes 1994; Achtermann \& Lacy 1995; Satyapal \etal
1997; Engelbracht \etal 1998; see also Vanzi \& Rieke 1997).  Studying
a sufficiently large sample of starbursts is therefore valuable in
breaking the degeneracy between age and upper mass cutoff.  As we are
not likely to catch all starbursts at a ``late'' age, and the
ultraviolet, optical, and far-infrared luminosities decline with time
after the peak star formation activity, we can determine the relative
importance of variations in star formation parameters and aging
effects by studying a larger sample.

In this work, we present results of a starburst modeling program to
interpret mid-infrared (MIR) spectroscopy from the Short Wavelength
Spectrometer (SWS; de Graauw \etal 1996) aboard the Infrared Space
Observatory (ISO; Kessler \etal 1996).  To this end, we have gathered
atomic fine structure line fluxes (in particular for the [Ne~II]
12.8$\mu$m and [Ne~III] 15.6$\mu$m lines) for 27 starburst
galaxies.  This project is ideally suited for investigating the
properties of the high-mass stellar population in heavily
obscured star-forming regions of starburst galaxies, since the
extinction at MIR wavelengths is only a few percent of the optical
extinction.  This investigation thus constitutes an important
contribution to the understanding of star formation in starburst
systems, a field that has been dominated by studies in the optical and
ultraviolet regimes in the intervening years between the IRAS and ISO
missions.  As recent studies of the cosmic infrared background suggest
a significant contribution from dusty starburst systems (e.g., Puget
\etal 1996; Hughes \etal 1998; Hauser \etal 1998; Elbaz \etal 1998),
the understanding of dusty starbursts is important in defining the
picture of the most powerful star-forming events in the universe.

We present the sample data in \S\ref{Sdata}. We outline our modeling
procedure and compare our ISO-SWS observations with model predictions
in \S\ref{Smodels}. We examine additional constraints on this
starburst modeling study and examine the robustness of the modeling
analysis in \S\ref{Sanalysis}. We discuss the implications of our
findings and summarize this work in \S\ref{Ssummary}.

\section{Neon Ratios in a Starburst Sample} \label{Sdata}

\subsection{Database selection}

The database for this project was created by combining data from the
MPE guaranteed-time program on bright galactic nuclei (15 objects) and
an open-time observing proposal (12 objects), resulting in an ISO-SWS
spectroscopic dataset for a total sample of 27 galaxies.  The objects
from the guaranteed-time program are all well-studied, infrared-bright
starburst galaxies, and the open-time sample objects were selected on
the basis of their 60\um~flux densities (S$_{60\mu m}$ $>$20~Jy) and their
observability by ISO in the last few months of the mission.  

Within these programs, we have targeted observations at a specific set
of spectral lines, particularly the [Ne~II] 12.8\um~and [Ne~III]
15.6\um~fine structure lines; in this work we will concentrate on the
ratio of these two lines as a tracer of the massive star population,
with additional lines introduced to constrain metallicities and gas
densities.  The neutral Ne atom and the $\rm Ne^{+}$ ion have
ionization potentials of 21.56 eV and 40.95 eV respectively, making
the [\ion{Ne}{3}]/[\ion{Ne}{2}] line ratio very sensitive to the
spectral shape of the UV radiation field and thus to the properties of
the most massive stars present. This information can be used to
provide constraints on various star formation parameters, including
the upper mass cutoff of the IMF and the age and duration of the
starburst (see Kunze \etal 1996 and Rigopoulou \etal 1996 for earlier
models using this and other MIR line ratios).

We have chosen the neon line ratio as the focus of this larger study,
both for its sensitivity to massive, hot stars as well as the minimal
uncertainties introduced in constructing the ratio of these two lines
from ISO-SWS observations.  The [\ion{Ne}{2}] and [\ion{Ne}{3}] lines
span a wider range in ionization potential than other line ratios in
this wavelength region, such as
[\ion{S}{4}](10.5\um)/[\ion{Ne}{2}](12.8\um),
[\ion{S}{4}]/[\ion{Ar}{3}](9.0\um), [\ion{Ne}{2}]/[\ion{Ar}{3}](e.g.,
Achtermann \& Lacy 1995; Beck, Kelly, \& Lacy 1997; Engelbracht \etal
1998), [\ion{Ar}{3}](9.0\um)/[\ion{Ar}{2}](7.0\um), and
[\ion{S}{4}](10.5\um)/[\ion{S}{3}](18.7\um)(e.g., Kunze \etal 1996;
Rigopoulou \etal 1996). In addition, the [\ion{S}{4}]/[\ion{Ne}{2}]
ratio has inherent uncertainties due to the determination of
metallicities, and all of the alternatives listed above are subject to
corrections for aperture size variations and/or systematic
uncertainties due to measurements in different ISO-SWS bandpasses.
Systematic effects are minimized for the neon lines because they were
observed with the same ISO aperture and bandpass. In addition, because
the wavelengths of the two MIR neon lines are close, extinction
effects can be neglected.  For example, for a dust distribution in M82
which is well mixed with the stars and obeys a standard extinction law
(e.g. Draine 1989) with A$_V\sim$50 mag (McLeod \etal 1993; Schreiber
1999), the observed line ratio is higher than the intrinsic line ratio
by only a few percent.  The same is true in the ultraluminous galaxy
Arp 220 for a uniform foreground dust screen providing A$_V\sim$50 mag
for the starburst emission (Sturm \etal 1996).  By contrast,
extinction corrections can change line ratios such as
[\ion{Ar}{3}]/[\ion{Ar}{2}] and [\ion{S}{4}]/[\ion{S}{3}] by 30-60\%
in similar conditions.  Furthermore, the [\ion{Ar}{3}] and
[\ion{S}{4}] lines lie in the same spectral region as a broad silicate
absorption feature at 9.8\um, further complicating the measurement of
line fluxes.

\subsection{Observations and Data Reduction}\label{SSobs}

The majority of the data described herein was taken in grating line
scan mode SWS02, with data from some galaxies taken from
full-grating-scan SWS01 observations.  Data reduction was completed
using the ISO SWS Interactive Analysis (IA) package, which includes
interactive tools for dark current subtraction, flat fielding, and
removal of fringing caused by interference within the detectors
themselves (see e.g., Schaeidt \etal 1996). The reduction completed

for this paper used post-mission calibration files (June 1998).  

Line fluxes were determined by integration over the observed line
profiles, which were generally close to Gaussian in shape.  The
systematic calibration of line fluxes has an uncertainty of $\le$30\%
(Schaeidt \etal 1996), though the effect of this uncertainty is
minimized by comparing two lines from the same bandpass.  The
uncertainties in the measured fluxes range from ~$\le$10\% for strong
line sources to 30-40\% for weaker sources (generally from the
open-time proposal).  For comparison with models (\S\ref{Sanalysis}),
we will assume uncertainties in the line ratios for the entire sample
to be $\pm$50\%.  The aperture size for the bandpass containing the
[Ne~II] and [Ne~III] lines is 14\arcsec~x~27\arcsec.  Because of the
weak sensitivity of this line ratio to extinction, no corrections for
extinction were made (though see \S \ref{SSparams} for other effects
of dust).

Table \ref{Tlines} lists the MIR neon line strengths for the galaxy
sample. Figure \ref{Flines} shows the corresponding neon line spectra,
and Figure \ref{Flirvsratio} shows the range of observed
[Ne~III]/[Ne~II] ratios as a function of the infrared luminosity.  The
infrared luminosity $L_{\rm IR}$ is computed from the {\em IRAS} flux
densities following the prescription of Sanders \& Mirabel (1996,
their Table 1): $L_{IR} = L_{\rm 8 - 1000~\mu m} = 5.6 \times
10^{5}\,D_{\rm Mpc}^{2}\, (S_{100} + 2.58S_{60} + 5.16S_{25} +
13.48S_{12})$, where $S_{\lambda}$ is the flux density in Jy in the
band centered at wavelength $\lambda$.  For several of the galaxies,
the sizes of the emission regions exceed the SWS field of view, so we
have applied a aperture-size correction to $L_{IR}$ based on the spatial distribution
of MIR or radio continuum emission, millimetric CO emission, or
near-infrared hydrogen recombination line emission from the
literature, assuming they all trace the same emission regions as the
MIR neon lines.  For eight galaxies, the observed $L_{\rm IR}$ was
multiplied by a factor less than 0.5 to correct for the smaller SWS
field of view, and for two galaxies there was insufficient data to
determine an aperture-size correction.  For the remaining galaxies, 11
required no correction, and seven galaxies were corrected by factors
between  0.5 and 1.0.

In general, all of the sample objects have neon ratios between 0.05
and 1.0; however, the low-metallicity systems NGC~5253 and IIZw40 have
neon ratios of 3.5 and 12, respectively.  In \S\ref{SSparams}, we
discuss the effects of metallicity on model line ratios, in order to
address the observed range of emission properties for the entire
sample.  The sample presented in this paper includes nine objects for
which [Ne II] line fluxes were measured by Roche \etal (1991), who
took observations from the ground in the 8-13\um~atmospheric window.
It is difficult to compare the fluxes derived from these two sets of
observations due to differences in aperture size and the uncertain
contribution of the broad 12.7\um~PAH feature at the end of the
8-13\um~band. This is particularly true for galaxies such as M82 which
exhibit strong PAH features (e.g., Roche \etal 1991; Sturm \etal
2000); in the case of M82, we estimate from the SWS data that the
broad 12.7\um~feature could contribute a factor of 50\% more flux to
the apparent [Ne~II] line at the resolution observed by Roche
\etal. In addition, for some objects in the Roche \etal sample the [Ne
II] line is redshifted out of the 8-13\um~wavelength range and so
cannot be measured from the ground.  However, for the 7 objects for
which comparisons could be made, the line fluxes are consistent to
within these rather large uncertainties.  NGC~5253 and IIZw40 were
also singled out by Roche \etal (1991) as displaying high-excitation
spectra, though the [Ne II] line was not detected. Roche \etal noted
that these galaxies had little emission from broad PAH features, which
is consistent with high excitation and UV energy density in a
low-metallicity environment (see also Vigroux 1997; Thuan \etal~1999;
and Madden 2000).

For further comparison, we include in Figure \ref{Flirvsratio} the
neon ratios for similar observations of W51, 30 Doradus, and the
central parsec of the Galaxy.  The neon ratios for W51 and 30 Doradus
are higher than for the starburst galaxy sample, while that for the
Galactic Center is lower (a further discussion of the Galactic Center
can be found in \S \ref{SSgc}).  Our ISO-SWS data thus generally
confirm findings from NIR and MIR spectroscopy that powerful dusty
starbursts exhibit low nebular excitation in comparison to Galactic
HII regions (e.g., Doherty \etal 1994; Cox \etal 1999).  However, we
find a range of values, with the most excited regions having neon
ratios similar to those of young, Galactic HII regions.

\section{Nebular emission from young stellar clusters} \label{Smodels}

The starburst modeling method presented here predicts MIR spectral
line ratios for evolving stellar clusters.  Observations of nearby
starburst galaxies reveal that starburst regions are in fact composed
of many individual units.  In addition to the HST detections of super
star clusters in starburst galaxies, bright compact sources seen in
high-resolution near-infrared broad-band images have been interpreted
as young clusters of red supergiants ({\em e.g.}  Tacconi-Garman,
Sternberg \& Eckart 1996; Satyapal \etal 1997). Furthermore, important
substructure in the ISM on 10-parsec- or even parsec-scales is
observed or inferred from modeling ({\em e.g.} Carral \etal 1994; Shen
\& Lo 1995; Achtermann \& Lacy 1995).

In \S\ref{SevolLF}, we present a set of models in which the
starburst region is composed of a heterogeneous ensemble of
recently-formed star clusters.  Therefore, we first lay the groundwork
of modeling the emission from the HII region around stellar clusters.

\subsection{The SED produced by a stellar cluster}

We have used the evolutionary synthesis code STARS (Sternberg 1998) to
model the integrated properties of individual stellar clusters.  This
code is based on the most recent Geneva stellar evolutionary tracks
(Schaller \etal 1992; Schaerer \etal 1993a, 1993b; Charbonnel \etal
1993; Meynet \etal 1994), and is similar to other models developed by
various investigators ({\em e.g.}  Tinsley 1972; Huchra 1977; Bruzual
1983; Guiderdoni \& Rocca-Volmerange 1987; Mas-Hesse \& Kunth 1991;
Bruzual \& Charlot 1993; Leitherer \& Heckman 1995; Leitherer \etal
1999).  STARS follows the evolution in the H-R diagram (or,
alternatively, in a $\log T_{\rm eff} - \log g$ diagram) of a stellar
population whose composition is specified by a time-independent IMF.
The star formation rate (SFR) is assumed to decline exponentially as
$e^{-\tburst/\tsc}$, where \tburst~ is the age of the cluster and
\tsc~ is the burst timescale.  At a given age, the integrated SED is
obtained by summing over the contributions of all stars present.

The SED assigned to each location in the $\log T_{\rm eff} - \log g$
diagram is taken from a hybrid grid created from two separate
libraries, in order to best represent the properties of all stars.
For very hot stars, the effects of line blanketing and blocking in
rapidly expanding atmospheres are not well represented by LTE models;
thus for stars with temperatures above 25,000~K (M$_{\rm ZAMS}\ga$
13\msolar), we have used new, non-LTE SEDs by Pauldrach \etal (1998).
These models include the effects of stellar winds and mass loss, and
cover effective temperatures up to 60,000~K (M$_{\rm ZAMS}\sim$
120\msolar).  The ``cornerstone'' grid of Pauldrach models consists of
8 main sequence models and 6 supergiant models, from which SEDs for
intermediate values of Teff and log g are interpolated.  Table
\ref{Torigpaul} lists the (T$_{eff}$, log g) combinations in the
cornerstone grid, and Figure \ref{Fsedgrid} shows the sampling of the
interpolated grid of solar-metallicity models.  For stars with
effective temperatures below 19,000~K (M$_{\rm ZAMS}<$ 8\msolar), we
use the Kurucz (1992) library of LTE SEDs.  ``Hybrid'' SEDs for stars
with intermediate effective temperatures were created by interpolating
between the 19,000~K Kurucz model and the 25,000~K Pauldrach
model. Figure \ref{Fseds} compares examples of Kurucz, Pauldrach, and
hybrid SEDs over a range of temperatures.

We have created SED grids for two metallicities, solar (\zsolar) and
0.2\zsolar. The low-metallicity grid allows us to better represent the
stellar populations in NGC~5253 ($\sim$0.24\zsolar, Hunter, Gallagher,
\& Rautenkranz 1982) and IIZw40 ($\sim$0.15\zsolar, Masegosa, Moles,
\& Campos-Aguilar 1994), and provides a benchmark for the effects of
metallicity variations on our models.  The low-metallicity grid
differs from the solar-metallicity grid only in the sampling of the
cooler Kurucz models. Starburst models were computed using each grid.

Table \ref{Tmodelpars} summarizes the model parameters used in this
work.  For each metallicity grid, only the upper mass cutoff (\mup),
burst age (\tburst), and burst timescale (\tsc) were allowed to
vary. The IMF was taken to have a Salpeter power-law index, ${\rm
d}N/{\rm d}m \propto m^{-2.35}$ (Salpeter 1955), between a lower mass
cutoff (\mlow) fixed at $1~{\rm M_{\odot}}$ and a variable \mup.
Though some studies suggest that the IMF in starburst galaxies is
deficient in stars below a few solar masses ({\em e.g.}  Rieke \etal
1980, 1993; Shier, Rieke \& Rieke 1996; Engelbracht \etal 1998),
variations in the low-mass IMF do not affect the ionizing continuum
and the results presented here; thus we will not consider variations
in \mlow.

\subsection{Modeling the nebular emission around clusters}\label{SSsingcl}

To model the nebular emission excited by a stellar cluster, we have
used the photoionization code CLOUDY (version C90.04, Ferland 1996),
with the source of ionizing radiation described by SEDs from STARS 
and the ionized nebula represented by a shell around a
central stellar cluster.  The nebular emission from young clusters
will depend on the properties of the surrounding ISM.  The nebular
conditions are specified by the gas and dust composition, the hydrogen
number density $n_{\rm H}$ of the gas, the distance $R$ between the
ionizing source and the illuminated surface of the cloud, and the
ionization parameter $U$.  To estimate the appropriate inputs for
these models, we have collected measurements from our data and from
the literature which can be used to infer metallicities, $n_{\rm H}$,
$R$, and $U$ for as many sample objects as possible.

Using our neon line measurements and published hydrogen recombination
line measurements, we have calculated the abundance of neon relative
to hydrogen for the sample using the strong-line method.  This method
provides abundances good to within factors of 2-3 (Osterbrock 1989),
sufficient for our purposes here.  Excluding NGC~5253 and IIZw40, the
calculated metallicities are consistent with solar metallicity, to
within the uncertainties: the average metallicity for the 13 objects
for which metallicities could be determined was 1.9$\pm$1~\zsolar.
The metallicities of these objects will be more rigorously derived by
Spoon \etal (in prep).  

We have computed the photoionization models assuming a solar
gas-phase abundance for the majority of the sample, as well as models
assuming a 0.2~\zsolar~gas-phase abundance for NGC~5253 and
IIZw40.  As changes in gas-phase abundances have relatively little
effect on the neon ratio, we have not attempted to further tailor the
gas-phase abundances for each galaxy.  The comparison between solar
and sub-solar metallicity models will be discussed further in \S
\ref{SSparams}.

Due to uncertainties about the dust properties in the starburst
environments we have surveyed, we neglect the effects of dust grains
mixed with the ionized gas.  The gas density is assumed to be uniform,
and we take $n_{\rm H} = n_{\rm e}$, where $n_{\rm e}$ is the electron
density.  The electron density was determined from the [\ion{S}{3}]
18.7\um/33.5\um~line ratio for the galaxies in the guaranteed-time
sample, with the line fluxes corrected for the extinction determined
in \S \ref{SSbollyc}, assuming the extinction law recommended by
Draine (1989).  The typical uncertainties on the line ratios are 50\%,
and are dominated by extinction and aperture-size corrections (the
[\ion{S}{3}] 18.7\um\ and 33.5\um\ lines were measured through
$14^{\prime\prime}\times27^{\prime\prime}$ and
$20^{\prime\prime}\times33^{\prime\prime}$ apertures, respectively),
and the absolute flux calibration.  Within the uncertainties, the
dereddened line ratios for this subsample lie in the low-density
limit, and imply $n_{\rm e} \la 10^{3}~{\rm cm^{-3}}$.  We adopt a
value of $300~{\rm cm^{-3}}$ for all models, inferred from the average
of the line ratios for the galaxies in the guaranteed-time sample.

The ionization parameter (U) of a nebula is one of the
most important parameters in photoionization models.  A measure of $U$
gives the number of ionizing photons impinging at the surface of the
nebula per hydrogen atom:
$$
U \equiv \frac{Q_{Lyc}}{4 \pi R^{2}n_{\rm H}\,c},
\eqno(1)
$$
where $Q_{Lyc}$ is the rate of production of ionizing photons, $R$ is
the radius of the shell, and $c$ is the speed of light.  The
determination of the ionization parameter $U$ from observed properties
depends on the assumptions made about the geometry of the ionizing
clusters and of the nebulae.  As discussed above, entire starburst
regions such as included in the SWS field of view likely do not
consist of a gas shell illuminated by a single, centrally-concentrated
cluster. If the ISM and a number of young star clusters are in a mixed
distribution, gas clouds will shield each other partially from
ionizing radiation from clusters distributed throughout a starburst
region of radius R (e.g., Wolfire, Tielens, \& Hollenbach 1990). Both
representative geometries are shown schematically in Figure
\ref{Fgeom}.  In the distributed cluster geometry, we can describe the
conditions at the illuminated face of each cloud structure in terms of
an effective ionization parameter, U$_{\rm eff}$ which may be written

$$ \log U_{\rm eff} = \log \left(\frac{Q_{Lyc}}{4 \pi R^{2} n_{\rm H} c}\right)
+ \log \left\{\left(\frac{\lambda}{R}\right)\,
\left[1 - e^{-R/\lambda}\right]\right\},
\eqno(2)$$

The distributed cluster geometry is analogous to that in ``mixed
extinction'' models, where R/$\lambda$ is a measure of the optical
depth and $\lambda$ is the mean free path of photons. The parameter
$\lambda$ depends on the number density and sizes of the clouds which
reside in the starburst region.  Therefore, the determination of
$U_{\rm eff}$, and the corresponding effective radius $R_{\rm eff}$ (from replacing U
with U$_{\rm eff}$ in Eqn. 1), requires a knowledge of the relative
distributions of gas clouds and ionizing clusters, as well as the
properties of the clouds and clusters themselves.  Such information is
available for the archetypal starburst galaxy M\,82: by combining new
near- and mid-infrared data with the results of modeling of
photodissociation regions by Lord \etal (1996) and observations of the
molecular gas by Shen \& Lo (1995), Schreiber (1999) reassessed the
geometry of clusters and gas clouds in the starburst region of M82.
The effective values, from an investigation of the ionization
parameter in regions of varying size within the starburst core of
M\,82, are $\log U_{\rm eff} = -2.3$ and $R_{\rm eff} = 25~{\rm pc}$
(Schreiber 1999).

The investigation of M\,82 by Schreiber (1999) further shows that the
effective ionization parameter is essentially independent of the
location and the size of the regions studied, from individual burst
sites $\approx 20~{\rm pc}$ in radius to the entire starburst core
extending over $450~{\rm pc}$.  This strongly suggests that the
regions of intense starburst activity in M\,82, which dominate the
nebular line emission, have very similar properties over a wide range
of physical sizes.  On average, the nebular conditions of the gas in
the starburst regions of M82 can thus be described locally and globally with
the same parameters.

A similar analysis is possible for the entire starburst regions of
NGC\,3256 and NGC\,253.  By combining published data on the integrated
Lyman continuum luminosity from Rigopoulou \etal (1996) and
Engelbracht \etal (1998) with the results from PDR modeling by Carral
\etal (1994), we infer $\log U_{\rm eff} = -2.3$ and $-2.6$,
respectively.  The value of $\log U_{\rm eff}$ we derive for NGC 253
is consistent with that derived by Engelbracht \etal 1998 (log
U$\sim -2.2$ to $-2.5$).  The determination of log U for these three galaxies
supports the suggestion by Carral \etal (1994) that the ISM properties
in starburst galaxies are independent of the global luminosity and
the triggering mechanism, and are endemic to starburst activity
itself.  We therefore use the $\log U_{\rm eff}$ and $R_{\rm eff}$
determined in M\,82 as representative values for the entire sample.

In our models, $n_{\rm e}$ and $R = R_{\rm eff}$ are assumed to be
time-independent.  We first computed the neon line ratio for a very
short-lived starburst, with \tsc=1 Myr.  In this model, $U$ varies
proportionally with Q\slyc~over time, via Equation (1). To set the
absolute flux scale of the cluster SEDs, we choose the condition that
when each cluster reaches its maximum Q\slyc, $U_{\rm max} =U_{\rm
eff}$; the shape and {\em relative} flux scale as the
clusters evolve are determined from the composite SEDs output by
STARS.  The computations are stopped when $\log U$ becomes smaller
than $-5.5$; at this point, the SFR has dropped down by several orders
of magnitude compared to the initial SFR, and the stellar population
produced by the starburst is expected to have faded away into the
galaxy's background population.

With the nebular parameters determined above as constraints, the SEDs
produced by STARS are input into CLOUDY.  Approximating a thin shell
around the distributed cluster geometry, CLOUDY was run with
plane-parallel geometry and an ionization parameter specified by
Equation (1).  Models for the burst timescales presented in this paper
were computed by convolving the neon fluxes for the \tsc=1 Myr
burst with an exponential SFR with the appropriate value of \tsc,

$$
F_\lambda(t_b) = \int_{0}^{t_b}F_\lambda(t_{sc}=1 Myr,\tau) e^{-(t_b -
\tau)/t_{sc}} d(\tau) ,
\eqno(3)
$$

where F$_\lambda$ is the flux in the line at wavelength $\lambda$.
This method inherently assumes that U(\tburst) is proportional to the
ionizing photon rate produced in the 1 Myr-timescale burst, rather
than that indicated by the integrated Q$_{Lyc}$ for the longer burst
timescales. Thus at later times, the [Ne III]/[Ne II] ratio will stay
constant, as the line emission will be produced only by the most
recently formed clusters.  It is these recently-formed clusters which
will dominate the ionizing radiation field at all ages, and are all
assumed to have the same $U_{\rm max}$. 

\subsection{Results for a homogeneous sample of evolving clusters}\label{SScluster}

The model neon line ratios output by CLOUDY are indicated by the solid
lines in the upper panel of Figure \ref{Fsbmods}, as a function of
burst age and upper mass cutoff for various \tsc.  These ratios assume
no dust, n$_e\sim$300~cm$^{-3}$, and log U$_{max}=-2.3$.  The model
predictions for five different upper mass cutoffs ($m_{\rm up}$=25,
30, 35, 50, and 100~$\rm M_{\odot}$) and three burst timescales
(\tsc=1, 5 and 20~Myr) are shown to represent a range of prescriptions
for the formation of stars in clusters.  There is a slight decline in
the computed ratio for \tburst$\sim$5~Myr for the longest timescales,
due to a build up of evolved clusters that soften the shape of the
integrated ionizing radiation field.  However, these evolved clusters
must still be relatively young to affect the integrated ratio
significantly.  The asymptotic behaviour of the curves for timescales
longer than $\approx 5~{\rm Myr}$ can be attributed mainly to our
choice of an exponentially decaying SFR, but also reflects the
dominance of the youngest generations, for which $\log U_{\rm eff}$ is
close to the maximum.

The vertical bar to the right of the plot represents the range of
observed neon line ratios for all sample galaxies except the two
low-metallicity blue dwarf galaxies NGC 5253 and IIZw40, which will be
discussed separately in \S\ref{SSparams}.  From the comparison with
models, values of $m_{\rm up}<$ $25~{\rm M_{\odot}}$ can be ruled out
for the entire sample.  In addition, the largest neon ratios in the
sample are consistent with upper mass cutoffs {\em at least as large}
as $50~{\rm M_{\odot}}$, independent of the burst age and timescale.
Figure \ref{Fsbmods} shows clearly the degeneracy between age and
\mup: the range of observed neon ratios may either be due to true
variations in the population of the most massive stars, or to short
bursts of star formation where \mup~is always large but the SED is
softened through aging.  However, a similar modeling analysis of the
nearby star formation sources plotted in Figure \ref{Flines} (Thornley
\etal, in prep.) and extensive modeling of M82 (Schreiber 1999)
support the need for stars more massive than 50\msolar, in agreement
with stellar census measurements in a variety of nearby star forming
regions.

\subsection{Modeling infrared emission from a heterogeneous ensemble of evolving clusters}\label{SevolLF}

The models described above are appropriate for individual clusters
(and surrounding nebulae) or for a homogeneous collection of clusters,
enclosed within a larger region.  However, the starburst galaxies we
are examining may contain evolving clusters with a wide range of total masses,
and low-mass clusters may not have enough material to fully sample the
IMF.  Here, we explore the effect of an assumed {\em cluster mass
spectrum} on the integrated [Ne~III]/[Ne~II] ratio.  On average, the
ionizing spectrum is softer if small clusters contribute a significant
fraction of the total ionizing luminosity.

Such an ensemble of clusters can be conveniently described by a luminosity
function (LF).  Various studies have explored the LF of young clusters
and HII regions in a wide range of star-forming environments,
including the disk of our own Galaxy (McKee \& Williams 1997), nearby
spiral and irregular galaxies ({\em e.g.} Kennicutt, Edgar \& Hodge
1989; Elson \& Fall 1985; Gonzalez-Delgado \etal 1995), as well as
merger systems like NGC\,4038/4039 (Whitmore \& Schweizer 1995).
These studies show that the number distribution of clusters in optical
light and in ionizing luminosity follows a power-law LF, ${{\rm
d}N}/{{\rm d}(\log L)} \propto L^{-\beta}$, with remarkably similar
indexes in the range $\beta = 0.5 - 1$, down to the faintest
luminosities observed.  The typical completeness limits correspond to
an absolute $V$-band magnitude of $\rm \approx -10~mag$ or to a Lyman
continuum luminosity of $\sim 10^{50}~{\rm s^{-1}}$.  Hence only
bright clusters are directly observed. For comparison, the Orion
nebula is powered by stars which emit a total of $10^{48.85}$ Lyman
continuum photons $~{\rm s^{-1}}$ (Kennicutt 1984).

In order to model the integrated emission properties of an
ensemble of clusters, we must include the contribution from all
clusters in a model starburst event.  While this calculation is
relatively straightforward in large, luminous clusters, it is more
uncertain in lower-mass clusters where statistical fluctuations at
the high-mass end of the IMF become important: for instance, the
formation of a single 20\msolar~ star instead of a single 15\msolar~
star could change the cluster luminosity by an order of magnitude. To
determine the shape of the LF consistently for the brightest clusters
as well as clusters fainter than $\sim$10$^{50}$~s$^{-1}$, we ran a
Monte Carlo simulation to determine the stellar contents of clusters
of different masses using Poisson statistics.

Our Monte Carlo simulation was run for a sample of 2x10$^5$ clusters,
with the number of stars per cluster ranging from 100 to 10$^{7.5}$
and mass bins populated from $m_{low}$=0.1\msolar~upward in accordance with
the specified Salpeter IMF. In this way, the upper mass cutoff of an
individual cluster is dependent on both the total mass of the cluster
and statistical variations in the population of the IMF.  Using
published values to represent the Q$_{Lyc}$(m) relation for individual
stars (Vacca, Garmany, \& Shull 1996; Panagia 1973), we then
determined the average cluster mass, m$^{\rm cl}$, as a function of ionizing
luminosity Q$_{Lyc}^{\rm cl}$ for the range of clusters falling in a
given luminosity bin of width 0.1 dex.  The resulting relation,
m$^{\rm cl}$(Q$_{Lyc}^{\rm cl}$), can be approximated by
two power laws, with m$^{\rm cl}\propto$ Q$_{Lyc}^{\rm cl}$ for clusters with
log Q$_{Lyc}^{\rm cl}>$49.5 and m$^{\rm cl}\propto$ (Q$_{Lyc}^{\rm
cl})^{0.19}$ for clusters with log Q$_{Lyc}^{\rm cl}<$49.5.  To
convert from a function m$^{\rm cl}$(Q$_{Lyc}^{\rm cl}$) to a
luminosity function, we assume a simple, power-law cluster {\em mass}
function of the form dN/d(log m) $\propto$ m$^{-\alpha}$, such that
dN/d(log Q$_{Lyc}^{\rm cl}$) $\propto$ (Q$_{Lyc}^{\rm
cl}$)$^{-\alpha\beta}$.

With these conditions, we derive a LF covering Q$_{Lyc}^{\rm cl}$ from
10$^{44}$ to 10$^{53}$ photons s$^{-1}$.  For the cluster mass
function, we assume the simple case $\alpha$=1, which produces a LF
shape consistent with that determined by McKee \& Williams (1997) for
HII regions in the disk of our own Galaxy.  The lower and upper limits
for Q$_{Lyc}^{\rm cl}$ in our LF were chosen to represent the smallest
associations still capable of ionizing an \ion{H}{2} region ({\em
i.e.} containing only one early-B star), and the most luminous super
star clusters detected ({\em e.g.,} in NGC\,4038/4039).  The resulting
LF is again represented by a broken power law, with $\beta$=0.19 at
the lower end and $\beta$=1.0, with the break occuring at
Q$_{Lyc}^{\rm cl}$ $\sim 10^{49.5}~{\rm s^{-1}}$ as for the m$^{\rm
cl}$(Q$_{Lyc}^{\rm cl}$) relation for individual clusters.  Figure
\ref{Flf} shows the derived LF, with the McKee \& Williams LF
overplotted for comparison.

We assume that the LF determined above describes the cluster ensemble
{\em at birth}, and then follow the evolution of the ensemble of
clusters for various star formation histories.  The evolution of the
LF with burst age for the ensemble of clusters is obtained by
following the evolution of Q$_{Lyc}^{\rm cl}$ for the clusters in each
luminosity bin.  For this purpose, we computed a library of
cluster models for a \tsc=1 Myr burst using STARS and CLOUDY.
The Monte Carlo simulations described above provide the relation
between Q$_{Lyc}^{\rm~cl}$ and $m_{\rm up}^{\rm cl}$, the mass of the
most massive star in an individual cluster given a fixed IMF and an
input stellar count (note that $m_{\rm up}^{\rm cl}$ should be
distinguished from an intrinsic, galaxy-wide upper mass cutoff
of the IMF, which we have thus far designated as $m_{\rm up}$).  

Models were generated for \tsc=1~Myr and $m_{\rm up}^{\rm
cl}$=5-100~${\rm M_{\odot}}$, in steps of $5~{\rm M_{\odot}}$. The
range of Q$_{Lyc}^{\rm cl}$ was divided in logarithmic bins of $\Delta
(\log $Q$_{Lyc}^{\rm cl}$) = 0.05~dex, and the models for intermediate
$m_{\rm up}^{\rm cl}$ were obtained by interpolation of the library
models. At zero age, the clusters with different initial $m_{\rm
up}^{\rm cl}$ (and Q$_{Lyc}^{\rm cl}$) were distributed in the
luminosity bins according to the LF.  As the burst age increases, the
evolution of each cluster in Q$_{Lyc}^{\rm cl}$ is followed (using the
library models), and the new LF at each time step is determined from
the distribution of the clusters in the different Q$_{Lyc}^{\rm cl}$ bins after
evolution has taken place.

The slope of the high-luminosity end of the LF ($\rm \ga 10^{50}~s^{-1}$)
changes very little during this evolution; this is mainly due to the
fact that these clusters contain stars with masses $>$50\msolar. All
such massive stars have similarly short main-sequence lifetimes ({\em
e.g.} Schaller \etal 1992), which implies that these luminous clusters
move into lower Q$_{Lyc}^{\rm cl}$ bins at similar rates.
The constant slope of the high-luminosity end of the LF is thus
consistent with the observed LF in various sources, which have
presumably a range of ages and starburst histories.

\subsection{Results for a heterogeneous ensemble of evolving clusters}\label{SSlf}

The neon ratio at any given age for the \tsc=1~Myr ensemble of
clusters was obtained by summing over the [Ne~II] and [Ne~III] line
fluxes of all the model clusters still contributing; the integrated
properties for longer burst timescales were again obtained by
convolving those for the \tsc=1 Myr burst.  The neon ratios predicted for
an ensemble of clusters distributed according to the derived cluster
LF are indicated, for the longest and shortest timescales we explored,
by the dashed lines in the upper panel of Figure \ref{Fsbmods}.  As
expected, the main effect of accounting for a cluster size
distribution is that the smaller, less luminous clusters soften the
integrated ultraviolet radiation field.  In particular, the predicted
[\ion{Ne}{3}]/[\ion{Ne}{2}] for an ensemble of clusters is lower than
for a single cluster, assuming the same galaxy-wide $m_{\rm up}$. The
effect of the LF is larger for higher \mup, with a reduction in
predicted ratios by a factor of $\sim$2 for a galaxy-wide
\mup=100~${\rm M_{\odot}}$, and $\sim$1.2 for \mup=25~${\rm
M_{\odot}}$.  This dependence on \mup~reflects primarily the stellar
properties themselves: the hardness of the ionizing spectrum decreases
steeply as the stellar mass decreases.  The magnitude of the effect of
incorporating a cluster LF depends mainly on the power-law index for
the LF and on the upper limit in Q$^{\rm cl}_{Lyc}$.  The steep
power-law we have adopted maximizes the differences between single and
ensemble cluster models, in comparison with other plausible LFs with
indices above the break of $\beta = 0.5 - 1$.  The choice of a high
upper limit in Q$^{\rm cl}_{Lyc}$ is justified by observations in some
starburst galaxies, notably NGC\,4038/4039 and M\,82 (Whitmore \&
Schweizer 1995; O'Connell \etal 1995).

Although the ``down-weighting'' effects of the smaller clusters, which
produce softer ionizing radiation, are measurable, they do not require
a significantly different conclusion than that reached by comparison
of our data with models of homogeneous clusters.  Accounting for a LF of the
ionizing clusters which excite the observed nebular emission lines, we 
confirm that very high-mass stars can form in starburst
galaxies, allowing $m_{\rm up} =50-100~{\rm M_{\odot}}$ even more
easily in the sources with the highest measured ratios.

\section{Additional considerations of the starburst scenario}\label{Sanalysis}

\subsection{Insights from the L$_{\rm bol}$/L$_{Lyc}$ ratio}\label{SSbollyc}

If we posit that all starbursts form stars in the manner that nearby,
massive star forming regions do, then the most plausible explanation
for low observed line ratios is that starbursts are events of short
duration (Figure \ref{Fsbmods} suggests \tsc$\sim$1~Myr) which produce
very massive stars, but whose aging rapidly softens diagnostic ratios
such as the ones we use here.  Our models show that as long as very
massive stars are formed, even in small numbers, they strongly
dominate the ionizing radiation field and thus maintain high neon line
ratios.  The ratios can only decrease to the observed range long
enough after the exhaustion of starburst activity for the most massive
stars to have evolved off the main sequence.  For ratios near 0.1,
corresponding to stars with initial masses near $30~{\rm M_{\odot}}$,
this will take about 5~Myr.  In principle, the neon ratios will start
declining very rapidly after the last massive stars have formed in the
burst, so that the neon ratio alone does not fully constrain the
timescale.  However, by combining with other measurements which are
indicative of starburst properties, we can explore the robustness of
our conclusion that starburst timescales are generally quite short. 

The ratio of the bolometric to Lyman continuum luminosities
(\lbol/\llyc) is a useful contrasting probe of the properties of
massive stars: since this ratio is sensitive to a somewhat lower mass
range than that to which \neonteff\ is sensitive, it varies
significantly with time even for longer burst timescales.  This
reflects the buildup of a population of stars in a lower mass range
which contribute more to \lbol, and which have longer main-sequence
lifetimes.  Therefore, the combination of both diagnostics can help
address the degeneracy between aging effects and variations in the
upper mass cutoff.

As we are characterizing the properties of massive star forming
regions, we assume $L_{\rm bol} = L_{\rm IR}$ (as defined in \S\ref{SSobs}).
This is usually a good approximation since a large fraction of the
energy output of OB stars is absorbed by the surrounding interstellar
dust, present in large amounts in starbursts, and re-emitted in the
thermal infrared. We derived L$_{Lyc}$ using two different methods.
First, we use measurements of hydrogen recombination line fluxes and thermal
radio continuum emission, when such data were available from
observations with the SWS and in the literature.  This allowed us to
constrain simultaneously the extinction toward the sources from the
comparison of observed and theoretical relative line fluxes.
Second, we used our own neon line fluxes corrected for extinction and
assuming all neon atoms are either singly- or doubly-ionized.

We assumed case B recombination coefficients and line emissivities
(Hummer \& Storey 1987), with an electron density of $n_{\rm e} =
100~{\rm cm^{-3}}$ and temperature of $T_{\rm e} = 5000~{\rm K}$,
except when individual determinations were available in the
literature.  For the extinction correction, we adopted a composite
extinction law made up of the Rieke \& Lebofsky (1985) law for
$\lambda \leq$ 0.9\um~ and the Draine (1989) law for 0.9\um $< \lambda
< $40\um.  The effects of obscuration were neglected for $\lambda >$
40\um.  If sufficient data were available, we constrained the geometry
of the emission sources and obscuring dust as well.  Two models were
considered: a uniform foreground screen (UFS) of dust and a
homogeneous mixture of dust and sources (MIX).  Otherwise, we
considered only the UFS model.

For the sources for which the determination from recombination lines was
possible, the estimates from the hydrogen and neon lines agree to within a
factor of three or better, and we adopted the average as the final
L$_{Lyc}$.  The cases where the extinction could not be reliably
constrained yield lower limits on L$_{Lyc}$, and thus upper limits on
L$_{IR}$/L$_{Lyc}$.  The lower panel of Figure \ref{Fsbmods} shows the
models for L$_{IR}$/L$_{Lyc}$ obtained with STARS.  As in the upper
panel, the ratios expected for a homogeneous cluster population are shown
as solid lines and those expected for a cluster ensemble defined by
our LF are shown as dashed lines. The effect of including a cluster LF
is of similar magnitude for \lir/\llyc~ as for the neon ratio.  It is
clear that the average neon and \lir/\llyc\ ratios are consistent with
conditions where the clusters have high upper mass cutoffs, $m_{\rm
up} \ga 50 - 100~{\rm M_{\odot}}$, and short burst timescales of a few
million to $\sim$10$^7$ years.  In fact, by plotting the neon ratios
against the \lir/\llyc\ ratios in Figure \ref{Fsbmodsllyc}, we see that
the models suggest very short timescales, so short as to be difficult
to produce even with the \tsc=1 Myr burst, the burst with the shortest
timescale considered in this study.  Though this discrepancy is
model-dependent, it is generally clear that short timescales are
needed to reproduce the range of ratios observed in our starburst
sample.

\subsection{The Galactic Center as a template for a short starburst}\label{SSgc}

The short timescales and ages inferred for the star forming activity
in starburst galaxies are reminiscent of the ones in the young stellar
clusters in the central parsec of our Galaxy, and can be checked there
on the basis of the existing stellar census. The observed
[Ne\,III]/[Ne\,II] ratio is even lower than in most starbursts,
despite the high ionization parameter log U = $-1$ derived for the
spatially resolved ionized region ($\la$~1~pc, Lutz \etal 1996). The
picture of a short but aged burst is supported by the direct stellar
census \citep{ght94,krabbe95,najarro97} which suggests a star
formation event of age $\sim$7 Myr and duration approximately
4~Myr. This event is most directly indicated by the presence of both
cool red supergiants like IRS7 and massive, moderately hot (20,000 to
30,000~K) blue helium-rich supergiants or Wolf-Rayet (WR) stars.
  
We have computed models optimized for this Galactic Center region,
assuming a single cluster with a Salpeter IMF, \tsc=1~Myr, \mup=100,
solar metallicity, and fixed log~U=$-1$, n=3000~cm$^{-3}$ as derived by
Lutz \etal (1996).  The burst timescale is somewhat shorter than that
suggested by the stellar census and was chosen as a conservative
assumption, since a shorter burst will produce softer radiation fields
at late ages and thus minimize any need to invoke other effects for
explaining the observed soft radiation.  The decay of line ratios with
time is similar to Figure \ref{Fsbmods}, but the ratios at any given
time are higher due to the higher log U in this smaller region. At the
age of 7 Myr preferred by the stellar census, the [Ne\,III]/[Ne\,II]
ratio is still about 1 to 2, well above the value of 0.05 observed by
Lutz \etal (1996). The model reaches the observed value only after
more than 13 Myr, which is difficult to reconcile with the presence of
evolved massive stars approaching 100 M$_\odot$ (Najarro \etal 1997;
Ott, Eckart, \& Genzel 1999).  While aging effects push the Galactic
Center neon ratio more closely into agreement with models, they are
not enough to fully account for the low observed neon ratio.

The stellar census for the Galactic Center gives an independent view
through direct analysis of the contributions of different stellar
types to the ionizing continuum. Figure~\ref{Fgcmod} presents an
Hertzsprung-Russell diagram weighted by Lyman continuum luminosity,
derived from the same starburst model for an age of 7 Myr. It is
evident that most of the ionizing luminosity still originates in stars close
to the main sequence at log T= 4.5 to 4.6, with smaller contributions
by stars somewhat evolved towards lower temperatures and a population
of hot WR stars to the left of the main sequence.  This
is in stark contrast to the finding by Najarro \etal (1997) that 7 of
their stars, found at log T $<$ 4.5 and log L $>$ 5.75, contribute
half of the ionizing luminosity of the Galactic Center. The same
region holds less than 1\% of the ionizing luminosity for the
model in Figure~\ref{Fgcmod}.  At all other model ages, this fraction
does not exceed about 1\%.

The fact that the discrepancy between `hard' models and `soft'
observations is seen both in the photoionization modelling and the
stellar census leads us to the conclusion that this discrepancy is not
primarily due to uncertainties in the hot star SEDs used for the
photoionization models. Such uncertainties affect mainly the shape of
the ionizing flux and thus the fine structure line ratios, but much
less the total ionizing flux of the star and the census.  Another
possibility is that the adopted Geneva stellar evolutionary tracks
\cite{schaller92} do not predict the large number of 20,000~K to
30,000K supergiants seen in the Galactic Center, or that there exists
a mismatch between the stellar effective temperatures from
evolutionary tracks and those from stellar atmosphere models. The
corresponding region is populated by these tracks but only for short
intervals, possibly because massive, post-main-sequence stars rapidly
move off to hotter parts of the tracks. The disagreement between
tracks and observations may be related to the difficulty of defining
mass loss, atmospheres, and effective temperatures for late stages of
massive star evolution governed by strong winds.  Our conclusions are
unchanged when using evolutionary tracks with twice solar metallicity
\cite{schaller92} and higher mass loss \cite{meynet94}, which might be
adequate for the somewhat higher metallicity in the Galactic
Center. The latter tracks in fact increase the disagreement by adding
more very hot WR stars.  In fact, WR stars may contribute to the
uncertainties in the evolutionary tracks, as our knowledge of the
ionizing spectra of WR stars is uncertain.  Indeed, there is both
observational and theoretical evidence \cite{crowth99,hillmill98}
which suggests that the ionizing spectra of WR stars may be much
softer than commonly assumed.

The case of the Galactic Center has obvious implications for aging
starbursts with similar populations. If the current Geneva tracks,
interpreted in the fashion described above, indeed mispredict the
post-main-sequence evolution of massive stars by postulating very hot
stars where the Galactic Center shows us a cooler luminous population,
then starburst models will tend to predict too high values for ratios
like [Ne\,III]/[Ne\,II].  The resultant ``need'' to invoke low upper
mass cutoffs (or aging effects) to explain soft radiation fields may
be due to this effect.  Corrections for this effect would likely lower
the neon ratios, and thus relax the stringent conditions of short
burst time scales, leading to more plausible timescales of $\la$10$^7$
years.  While a careful study is needed to test the reliability of
current evolutionary tracks, this analysis of the Galactic Center
supports our general conclusion that starbursts are ``normal''
star-forming environments, in the sense that they produce very massive
stars just as local star forming regions do.
 
\subsection{The effects of parameter variations}\label{SSparams}

Starburst modeling is, by the very nature of the star formation
process, a multi-parameter problem. To assess the robustness of the
results we have presented, we now discuss the effects of varying the
possible input parameters of our model, considering variations in
starburst SEDs and nebular parameters.  The most significant changes
in a starburst SED will arise because of our choice of metallicity,
model SED libraries for individual stars, or the shape of the IMF.
Changes in nebular parameters have effects of similar magnitude as
those of changes in SEDs.  The most significant effects will be due to
variations in the ionization parameter, though variations in the dust
population as well as the density of the gas in these regions also
have minor effects.  For simplicity, these variations are illustrated
for single-cluster models with \mup=100\msolar~and \tsc=1~Myr in
Figure \ref{Fvariation}.

\subsubsection{Metallicity}
Of the above effects, those of metallicity are the most significant.
Metallicity effects are twofold: at sub-solar metallicities, the
stellar SEDs are harder due to reduced line blanketing and blocking
(e.g., Pauldrach \etal 1998), and the evolutionary track each star
follows through the H-R diagram changes, producing a ''hotter'' main
sequence.  Figure \ref{Fvariation}a compares the neon ratios for two
models, the hybrid SED grid at solar metallicity, and the
corresponding hybrid grid with stars and gas at 0.2~\zsolar, which is
more appropriate for the two low-metallicity dwarf galaxies in our
sample.  Lowering the metallicity results in considerably increased
neon ratios: for Z=0.2\zsolar, the predicted neon ratios increase by
factors of $\sim$4-10.  The increase in neon ratios at low metallicity is
dominated by changes in the evolutionary tracks and SEDs of model
stars, with the corresponding changes to gas-phase abundances playing
a relatively minor role.  NGC 5253 and IIZw40 have neon ratios 3 and
13 times higher, respectively, than in any of the other starbursts in
our sample; the observed ratio values are also plotted as horizontal
grey lines in Figure \ref{Fvariation}a.  Therefore, even accounting
for low metallicity, the neon ratios measured for these systems are
consistent with a stellar population with \mup$>$50-100\msolar. In
contrast with the example of Figure \ref{Fvariation}a, models with
Z$>\zsolar$ will have correspondingly lower predicted neon ratios;
this factor is worth considering if more accurate determinations of
abundances become available.

\subsubsection{SED libraries} 
If we were to choose Kurucz instead of Pauldrach SEDs to represent the
most massive stars, the contribution of the softer Kurucz spectra for
high-mass stars would produce lower neon ratios (cf. Figure
\ref{Fseds}). The significance of this effect can be seen in Figure
\ref{Fvariation}b, which shows the neon ratio for two input stellar
grids: the interpolated, hybrid grid we are using, and a standard
Kurucz model grid.  At a burst age of $\sim$3~Myr, the predicted neon
ratio using Kurucz SEDs is a factor of $\sim$2 lower than that
predicted by our hybrid grid. Comparison with our ISO-SWS data would
suggest \mup$>$50\msolar~ for more than half the sample, even at
zero-age.

\subsubsection{IMF slope}
We have taken a Salpeter IMF to be the most representative form of the
initial mass function; however, the universality of the IMF is still
debated (see, e.g., Massey 1998 and Scalo 1998 for contrasting views).
Some authors have suggested that the IMF in active star-forming
environments is well-represented by a Salpeter form (Hunter 1995;
Massey \& Hunter 1998). While the alternative model favored by
Scalo (1998) exhibits a steeper function at intermediate masses, it is
similar to Salpeter at the high-mass end (M$\ge$10\msolar), where the
neon ratio is most sensitive.  To illustrate the effects of changing
the IMF, we show the resulting neon ratios for a Miller \& Scalo
(1979) IMF in Figure \ref{Fvariation}b, where this IMF is represented
as a power law with index $-1.4$ for M=1-10\msolar~ and $-2.5$ for
M=10-100\msolar.  The Miller-Scalo IMF is generally steeper than
Salpeter, and thus there are fewer massive stars formed relative to a
given number of low-mass stars. The net result is a softer composite
SED and a prediction of lower neon line ratios.  The decrease
in the predicted neon ratio caused by changing the IMF in this way is
smaller than that caused by changing SED libraries.

\subsubsection{Ionization parameter}
Figure \ref{Fvariation}c shows the predicted neon ratios for two
 alternate values of the ionization parameter: log U$_{max}=-1.5$, the
 upper limit for U assuming that all clusters lie at the center of the
 starburst nebular emission region of M82 (Schreiber 1999), and log
 U$_{max}=-3.5$, a value more similar to that derived by Wang,
 Heckman, \& Lehnert (1997) for the diffuse ionized medium (DIM).  An
 ionization parameter of log U$_{max}=-1.5$ would result in predicted
 neon line ratios $\ga$3 times higher out to $\sim$7~Myr for
 \tsc=1~Myr.  Such an increase would imply upper mass cutoffs between
 25 and 50~\msolar~ for any bursts with \tsc$\ga$5~Myr, without
 accounting for any aging effects. This value of U$_{max}$ is a factor
 of $\sim$3 greater than the highest value which is consistent with
 the M82 analysis.  The effect of reducing the ionization parameter to
 log~U$_{max}=-3.5$ is larger, causing a decrease in the predicted
 neon ratios by an order of magnitude.  The highest observed ratios in
 our starburst sample would not be reproduceable, even for the most
 massive stars for which we have models, if log~U$_{max}=-3.5$.
 Previous starburst modeling studies along these same lines (see,
 e.g., Kunze \etal 1996, Rigopoulou \etal 1996) have used log
 U=$-2.5$, similar to the M82-based value used in this study.

\subsubsection{Dust population and gas density}
The presence of dust within HII regions can affect the efficiency of
nebular photoionization, but introducing a dust component has a
relatively small effect on our models.  Figure \ref{Fvariation}d shows
the effect of adding dust grains similar to those in Orion (we use the
Orion dust population that is incorporated in CLOUDY, from Baldwin
\etal 1991).  Adding such a dust component causes a $\sim$15\%
increase in the predicted line ratios.  This variation is smaller than
the uncertainties in the measured neon line ratios, and negligible
compared to the other parameter variations we have explored.  We
conclude therefore that we have introduced no significant
uncertainties by neglecting dust in earlier sections. Note, however,
that the uncertainties are larger for determinations of
L$_{\rm bol}$/L$_{Lyc}$, where dust could have a much more significant
effect.

Due to the large uncertainties in inferring the gas density from the
SIII (18.7/33.5\um) ratio, we also show in Figure \ref{Fvariation}d
the effects of increasing the gas density to 10$^3$ cm$^{-3}$.  This
is the highest density consistent with the range of measured sulfur
ratios and their uncertainties.  For the model shown, the neon line
ratio increases by less than five percent over the entire age range,
indicating that variations in gas densities have a negligible effect
on the output neon ratios.  However, we note that the ionization
parameter changes inversely with gas density, such that this increase
in gas density would also imply a drop in the ionization parameter to
approximately log U$sim-2.8$, thus solidifying the case for very
massive stars being present in the starbursts in our
sample.

\subsection{Extra-starburst contributions?}\label{SSother}

The models we have presented thus far assume that the only contribution to
the MIR neon line fluxes comes from direct photoionization by stars in
the starburst region itself.  In this section we examine constraints, from
ISO spectroscopy, on the possible contributions of
two other processes: excitation by active galactic nuclei (AGNs), and
contributions from a diffuse ionized medium (DIM) in the
surrounding galaxy (e.g., Lehnert \& Heckman 1994; Wang \etal 1997).

\subsubsection{\bf AGN contributions}
The MIR spectra of AGNs are distinctive in their display of strong
emission lines from highly ionized species, such as [Ne~V] and [O~IV],
which require higher excitation that can be produced even by the
hottest stars.  The absence of these lines, or their weakness relative
to lower excitation lines, has been used to demonstrate the dominant
contribution of star formation to the power produced in ultraluminous
IR galaxies (Lutz \etal 1996; Genzel \etal 1998).  Though some
starbursts show very weak [O~IV] emission, the most plausible
explanation is ionizing shocks from supernovae or superwinds (Lutz
\etal 1998).  The sample presented here significantly overlaps with
the Lutz \etal (1998) sample, and strong [O~IV] emission is generally
not seen; furthermore, the shock models which reproduce the weak
[O~IV] fluxes show negligible contributions to the [Ne~II] and
[Ne~III] lines analyzed here.  There are two possible exceptions.  In
NGC~6240, faint [O~IV] emission is detected but it is stronger than
that measured in ``normal'' starbursts. In NGC~7469, a comparison of
[O~IV] and [Ne~III] line profiles suggests some AGN contribution to
the [Ne~III] emission. we therefore consider the measured neon ratios
for NGC~6240 and NGC~7469 to be upper limits to the emission arising
from the starburst region.  With these exceptions in mind, we consider
the contribution from an AGN to be unlikely across the sample, and any
contribution must have a neglible effect on the results presented
here.
 
\subsubsection{\bf DIM contributions}
Studies of the Milky Way and other nearby galaxies suggest the
presence of a ``diffuse ionized medium'' (DIM), a gas component with a
relatively low ionization state and large scale height which permeates
the galaxy disk (e.g., Reynolds 1990; Dettmar 1992). It may be
difficult to exclude a DIM component as a contributor to the neon line
fluxes measured for this sample. Several studies have shown the
existence of diffuse, ionized emission in the disks of galaxies, which
is generally not associated with individual star forming regions.  For
the more distant objects in our sample, the aperture covers a large
physical area (the long axis of the SWS aperture corresponds to linear
diameters of 0.3-14~kpc for the galaxies observed), which may
encompass non-starburst emitting regions elsewhere in the galaxy.
Thus, we must consider the possibility that some form of DIM emission
may influence the integrated line fluxes. Lehnert \& Heckman (1994)
and Ferguson \etal (1996) showed evidence that the DIM may be produced
by ionizing starlight escaping from HII regions in the disks of
galaxies; Wang \etal (1997) measured its effects on large scale
measurements of optical excitation ratios such as [NII/H$\alpha$] and
[SII/H$\alpha$].  These studies suggest that the DIM may contribute as
much as 50\% of the global, integrated H$\alpha$ flux in spiral
galaxies.  If we assume that the DIM consists of ionizing radiation
escaping from young star clusters, with the same average parameters as
derived by Wang \etal (log U$\sim-4$, n$_e\sim$1 cm$^{-3}$), then the
model neon ratios could drop by a factor of two, making the effect of
a DIM contribution similar in magnitude to that of variations in the
ionization parameter that were discussed in \S \ref{SSparams}.

We cannot exclude the contribution from a low-density component, as
 the [S III] ratio generally provides only an upper limit to the
 density.  However, we have a qualitative constraint supplied by the
 range of neon line ratios observed in our sample: the observed neon
 ratios show no correlation with distance, as seen in Figure
 \ref{Fneondist}.  If the DIM were a significant contributor, we might
 expect the neon ratio to decrease with increasing source distance, as
 the aperture encloses an ever larger physical area. In the nearest
 galaxy in our sample, IC 342, the aperture encloses an area $\sim$200
 pc in diameter, and for galaxies at a distance of 30 Mpc the aperture
 still covers a region less than 5 kpc in diameter.  Thus, unless we
 are observing objects in which the size of the starburst area grows
 in proportion to its distance, we conclude that any DIM-like
 component in the sample galaxies does not make a significant
 contribution. Any DIM contributions will result in lower predicted
 neon ratios; thus, our conclusion that the most massive stars are
 generally formed in all starburst environments is not affected.

\section{Discussion and Summary}\label{Ssummary}

The cluster models that we have presented in this paper
support the formation of very massive stars (50-100\msolar) in
starburst galaxies.  While the quantitative estimate of \mup~for each
galaxy is model-dependent, it is clear that the formation of very
massive stars is necessary to explain the ionized line diagnostics
observed in this starburst sample.  This result suggests that while
starbursts produce prodigious amounts of energy and stars, the
high-mass stellar populations in starburst galaxies are not radically
different than those in high-mass star-forming regions observed
locally.  

\subsection{Short timescales for starburst activity}

As illustrated in Figures \ref{Fsbmods} and \ref{Fsbmodsllyc}, the
combination of the neon line ratio with the \lir/\llyc\ ratio strongly
favors the scenario for starburst activity where very massive stars
form, as in local smaller-scale starburst templates, and where the
burst last typically a few million to $\sim$10$^7$ years.  Such
timescales are shorter than previously thought (10$^{7} - 10^{8}~{\rm
yr}$; e.g., Thronson \& Telesco 1986; Heckman 1998).  It is clear that
detailed modeling is required to secure this result, including
additional constraints ({\em e.g.}  $K$-band luminosity, the rate of
supernova explosions, the depth of the near-infrared CO bandheads) and
spatially resolved information.  It is nonetheless in agreement with
other recent detailed studies of a few starburst galaxies, some of
which are also included in our sample ({\em e.g.} M\,82, Schreiber
1999; NGC\,253, Engelbracht \etal 1998).  As a result of instrumental
progress, there is now growing evidence that starburst activity occurs
in individual burst sites on physical scales of a few tens of parsecs
or less.  Short timescales are therefore naturally understandable
locally.  Our data, in conjunction with the other studies cited above,
provide evidence for short timescales on much larger scales,
suggesting that starburst activity also occurs {\em globally} on short
timescales, presumably as a result of one brief gas compression event,
or of successive episodes of such events separated by more than one
typical timescale. Short burst durations thus imply strong negative
feedback effects of starburst activity, globally as well as locally.

A simple argument can be invoked to explain the physical arguments
behind this result.  We can compare the cumulative kinetic energy
injected in the ISM by the supernova explosions over time ($E_{\rm
kin}$) with the binding energy of the gas ($E_{\rm grav}$), and assume
the starburst activity will stop when $E_{\rm kin}$ just balances
$E_{\rm grav}$.  This is a simplistic way of expressing the conditions
for a starburst wind to break out of the galaxy ({\em e.g.} Heckman,
Armus \& Miley 1990), but it is sufficient for order-of-magnitude estimates.

In order to relate $E_{\rm kin}$ to observed quantities,
we have considered the relationship between the rate
of supernova explosions \snrate\ and the bolometric luminosity \lbol.
Model predictions obtained with STARS for a variety of
star formation histories and upper mass cutoffs of the IMF
indicate that 
$$
10^{12}\,\left(\frac{\nu_{\rm SN}}{\rm yr^{-1}}\right)\,
\left(\frac{L_{\rm bol}}{\rm L_{\odot}}\right)^{-1} \sim 1
\eqno(4)
$$
as soon as the massive stars start to explode as supernovae, 
and as long as substantial star formation takes place.
It thus seems reasonable to assume that Equation (4) holds
for the bulk of the sample, likely having a range in age and
timescale but all exhibiting signs of significant,
recent starburst activity.

For simplicity, we here assume a spherical geometry for the
starbursts, with uniform mass distribution.  In addition,
we assume that each supernova explosion liberates
$E_{\rm mech}^{\rm SN} = 10^{51}~{\rm erg}$ of mechanical
energy, transferred as kinetic energy to the ISM with
an efficiency $\eta$.  The timescale $\tau$ for our
condition above satisfies :
$$
\eta\,\left(\frac{{\rm d}E_{\rm mech}^{\rm SN}}{{\rm d}t}\right)\,\tau \simeq
\frac{G\,M_{\rm dyn}^{2}}{R},
\eqno(5)
$$
where ${\rm d}E_{\rm mech}^{\rm SN}/{\rm d}t$ is the
rate of mechanical energy injection from the supernovae,
$G$ is the gravitational constant, $M_{\rm dyn}$ is the dynamical
mass of the system, and $R$ is the radius of the starburst
region.  Equation (5) can be re-written as
$$
\frac{\tau}{\rm Myr} \simeq 
\left(\frac{8.5}{\eta}\right)\,
\left(\frac{M_{\rm dyn}}{\rm 10^{9}\,M_{\odot}}\right)^{2}\,
\left(\frac{R}{\rm kpc}\right)^{-1}\,
\left(\frac{L_{\rm IR}}{\rm 10^{10}\,L_{\odot}}\right)^{-1}
\eqno(6)
$$
where we have substituted \lir\ for \lbol, appropriate for
our sample galaxies.
Application of Equation (6) to M82
($M_{\rm dyn} = 8 \times 10^{8}~{\rm M_{\odot}}$,
$R = 0.25~{\rm kpc}$,
$L_{\rm IR} = 4 \times 10^{10}~{\rm L_{\odot}}$), yields
$\tau \simeq 5\,\eta^{-1}~{\rm Myr}$, so for efficiencies
$\ga 10\%$, the estimated timescales are $\sim 10^{6} - 10^{7}~{\rm yr}$.

Our argument above is based on ``gas-disruption timescale'' estimates.
This differs from the conventional ``gas-consumption'' arguments,
which compare the star formation rates with the mass of the gas
reservoir.  In such estimates, the star formation rates are based on
comparison of absolute fluxes ({\em e.g.} H$\alpha$ fluxes, \lir) with
predictions from starburst models.  The estimates are thus very
sensitive to the assumed age and history of the starburst.  Our
estimates of the gas-disruption timescales are also model-dependent,
but have the advantage of being based on a quantity
($10^{12}\,\nu_{\rm SN}/L_{\rm bol}$) which varies by smaller factors.
Neither point of view accounts for further fueling processes, or dynamical
evolution of the systems ({\em e.g.}  starbursts in barred galaxies,
interacting/merging systems, etc.).  However, the
discussion presented here gives an alternative perspective to the
issue of global burst timescales, and provides a plausible explanation
for our results.

\subsection{Summary} 

Starburst models predicting the [Ne~III]/[Ne~II] ratio from ISO-SWS spectra
of 27 starburst galaxies show that the observed data are consistent
with the formation of very massive stars in starbursts,
thus precluding the need for the restrictive upper mass cutoffs
suggested by some earlier studies (\mup$\sim$25-30\msolar).  Combining
the neon line ratios with starburst modeling and the consideration
of the stellar content measured in local star forming regions, we find
that starburst events may be generally described as short bursts of
star formation which produce very massive stars, and which exhibit
relatively soft integrated line ratios as a result of aging the
stellar population.

In particular, our modeling of neon and \lir/\llyc\ ratios, together
with results on local high-mass star-forming regions, suggest:

$\bullet$ very massive stars (\mup$\ga$50\msolar) form in typical starbursts.

$\bullet$ starbursts have short global timescales, \tsc$\la$10$^7$
years.

These results suggest strong negative feedback from starburst
activity; the galactic superwinds frequently observed in starburst
galaxies are particularly striking examples of the consequences of
such feedback.

In our analysis, we have examined the degeneracy between aging effects
and model parameter variations in the assessment of upper mass cutoffs
to the IMF.  There is still room for significant improvements in
modeling the properties of starbursts: determination of metallicities
and the radiation environment (e.g., for measurements of the
ionization parameter U) compete with the characterization of stellar
properties (SEDs, evolutionary tracks) as the largest contributors to
uncertainty in the modeling of star formation properties such as the
upper mass cutoff.  Other datasets, such as additional MIR line ratios
(e.g., Kunze \etal 1996; Rigopoulou \etal 1996; Engelbracht \etal
1998) or K-band luminosities and near-infrared spectroscopy (e.g.,
Forbes \etal 1993; van der Werf \etal 1993; Genzel \etal 1995;
Tacconi-Garman \etal 1996; B\"oker, F\"orster-Schreiber, \& Genzel
1997; Engelbracht \etal 1998; Schreiber 1999) would be very useful in
further constraining the properties of starbursts. However, it will be
important to compile such additional data for a large sample in order
to proscribe further, {\em general} constraints on the way in which
starbursts form stars. Observations with higher spatial resolution
would better isolate regions of active star formation, making it
possible to confirm whether high- and low-excitation lines arise from
the same region; the spectroscopic capabilities of SIRTF will be
well-suited to addressing this issue.  By accounting for a reasonable
range of uncertainties which constrain the present observations, we
find that the observed MIR neon ratios are generally consistent
with the formation of very massive stars in starburst events; we offer
this hypothesis up to future datasets for increasingly rigorous
testing.

We would like to thank A. Pauldrach and R.-P. Kudritzki for providing
model atmospheres, Tal Alexander for assistance in introducing
low-metallicity SEDs into STARS, and Eckhard Sturm for providing the
ISO-SWS spectra of Arp 220.  We would also like to thank Jack
Gallimore and Dan Tran for interesting discussions.  MDT would like to
thank the Alexander von Humboldt-Stiftung and the NRAO\footnote{The
National Radio Astronomy Observatory is a facility of the National
Science Foundation operated under cooperative agreement by Associated
Universities, Inc.} for support during the production of this
work. This research also received support from the German-Israeli
Foundation under grant I-0551-186.07/97.  SWS and the ISO Spectrometer
Data Center at MPE are supported by DARA under grants 50 QI 8610 8 and
50 QI 9402 3.

\onecolumn

\epsscale{0.8}
\plotone{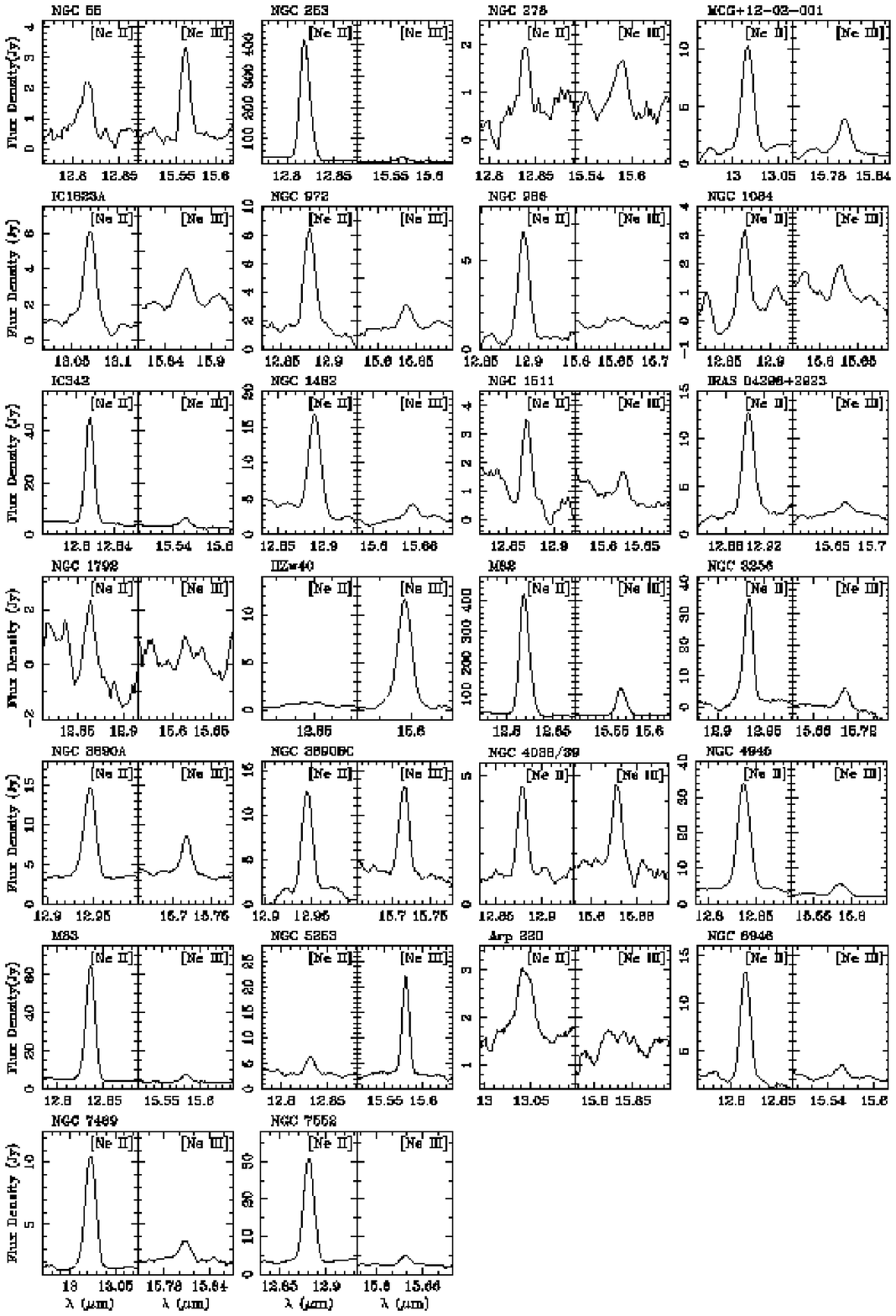}
\figcaption[f1.ps]{Spectra of the [Ne II] and [Ne III] lines
  observed in the sample galaxies. For each galaxy,
the [Ne II] and [Ne III] spectra are plotted on the same flux density scale
and over an identical velocity width. The spectra for NGC 6240 are
presented by Egami \etal (in prep.).\label{Flines}}

\epsscale{0.5}
\plotone{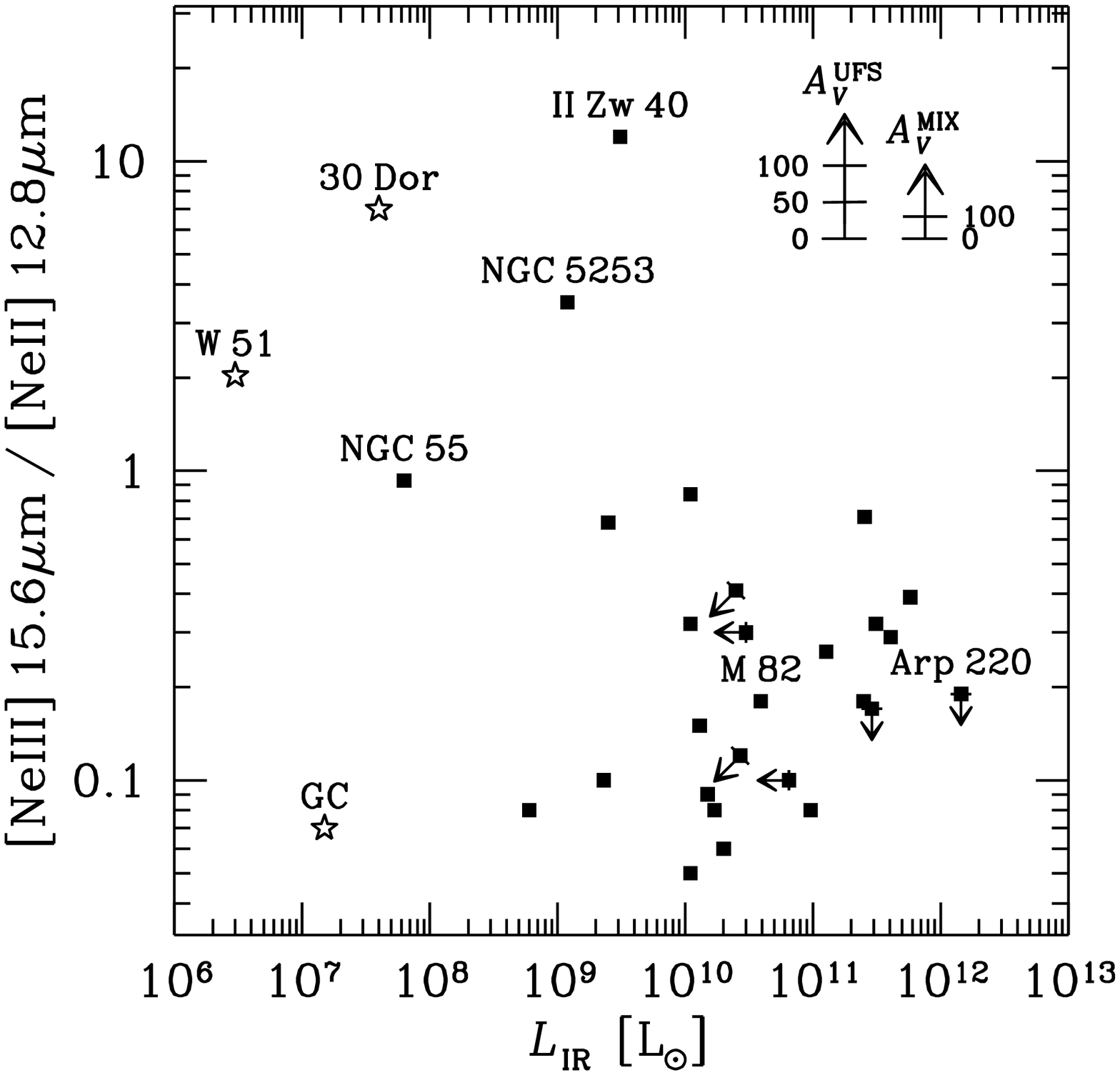}
\figcaption[f2.eps]{The [Ne III]/[Ne II] ratio as a function of
  infrared luminosity (L$_{\rm IR}$) for both the starburst galaxy
  sample (filled squares) and nearby ``template'' star-formation
  sources (open stars). The arrows in the upper right corner of the
  plot indicate the effects of extinction on the [Ne III]/[Ne II]
  ratio for two dust geometries: a uniform foreground screen of
  obscuring dust (``UFS'') and a homogenous mixture of the dust with
  radiation sources (``MIX'').\label{Flirvsratio}}

\plotone{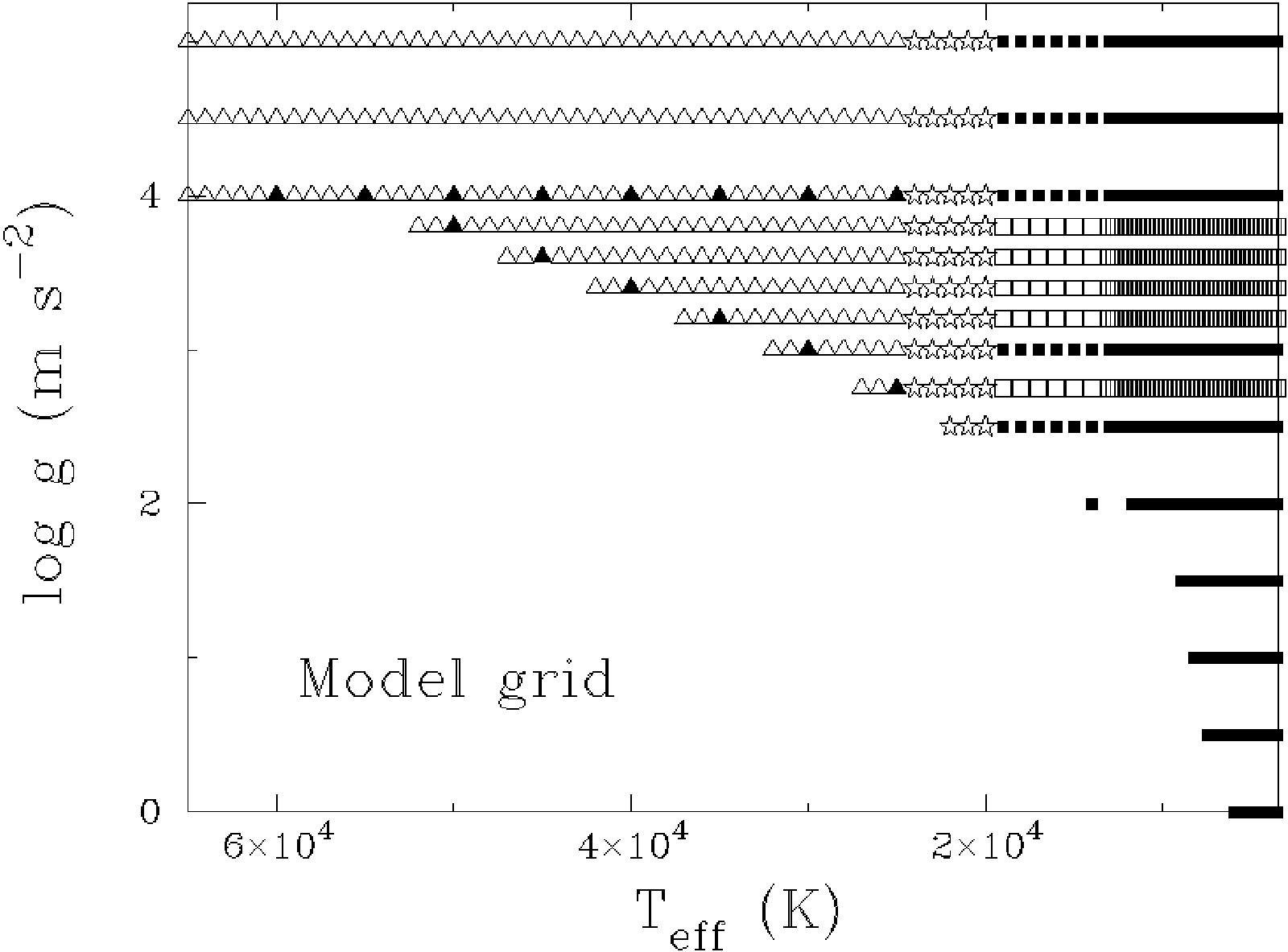}
\figcaption[f3.ps]{The interpolated grid of solar-metallicity
  model SEDs used in STARS to create composite SEDs. Squares indicate
  Kurucz models, triangles indicate Pauldrach models, and stars
  indicate ``hybrid'' models interpolated between Kurucz and Pauldrach
  models (see text). Original (``cornerstone'') models are shown as
  filled symbols, and interpolated and extrapolated models are shown
  as open symbols.  Models for log g$>$4 are identical to log g=4
  models, and models with T$_{eff}>$60,000~K are extrapolated from the
  highest temperature Pauldrach model.\label{Fsedgrid}}

\plotone{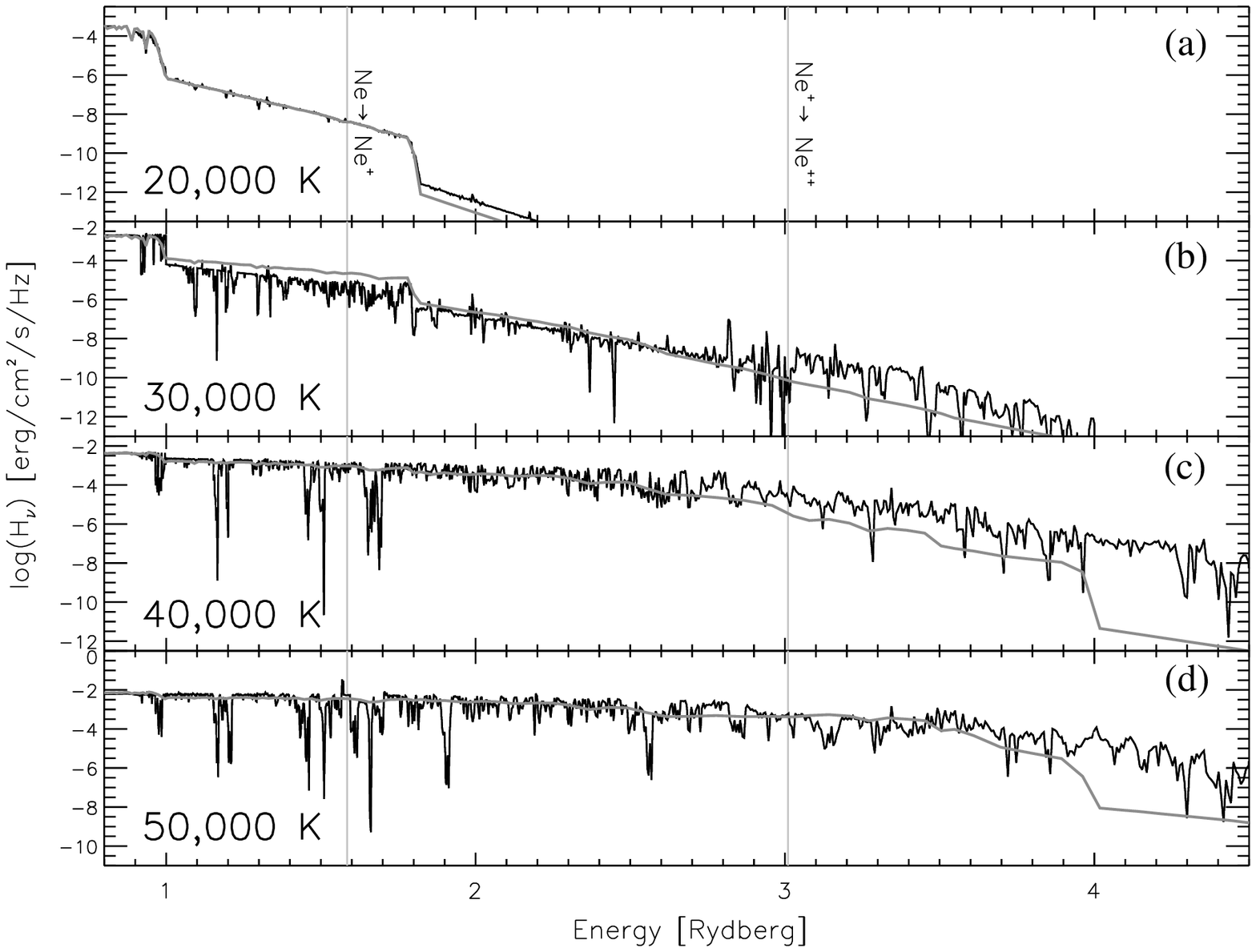}
\figcaption[f4.eps]{Examples of model SEDs in our interpolated,
solar-metallicity grid.  From top to bottom: {\bf(a)} Kurucz and ``hybrid''
models for a 20,000~K main sequence star (log g=4.0, corresponding to
a B3V star), where the Kurucz model is shown in grey and the hybrid
model in black.  {\bf(b)} Kurucz (grey) and Pauldrach (black) models for a
30,000~K main sequence star (B0V). {\bf(c)} Kurucz (grey) and Pauldrach
(black) models for 40,000 K and log g=4.5 (Kurucz) and 4.0
(Pauldrach). {\bf(d)} Kurucz (grey) and Pauldrach (black) models for
50,000~K and log g=5.0 (Kurucz) and 4.0 (Pauldrach).  In case (a), the
hybrid model is used; in all other cases, the Pauldrach model is
used.\label{Fseds}}

\epsscale{0.4}
\plotone{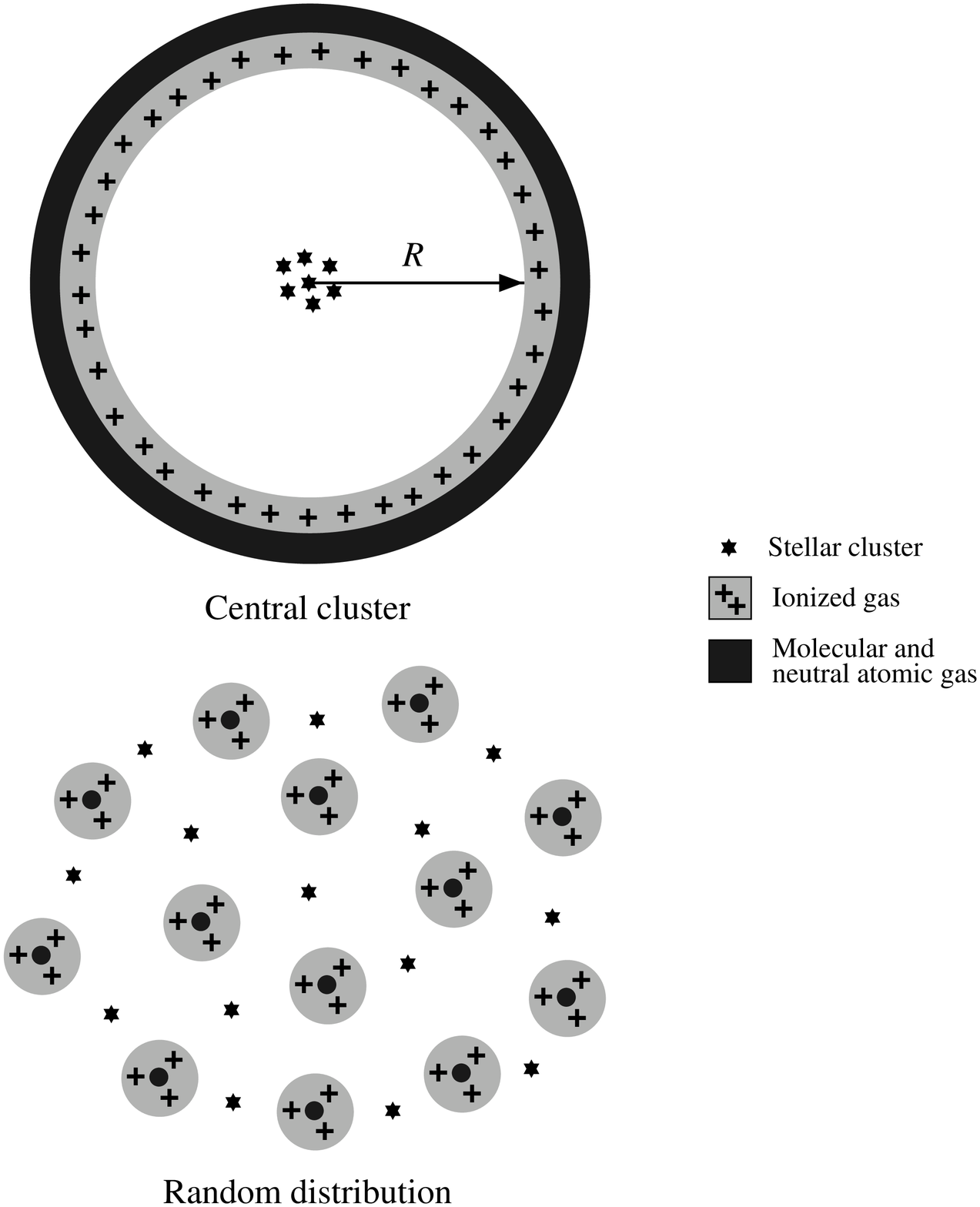}
\figcaption[f5.eps]{Two starburst region geometries, adapted from
Wolfire, Tielens, \& Hollenbach (1990): central cluster geometry
(top) and distributed cluster geometry (bottom).  The bottom picture
can be approximated by the top picture if the additional shielding of
the radiation field by the intermediate clouds is included in the
calculation of the appropriate ionization parameter U$_{\rm eff}$ (see
\S\ref{SSsingcl}).\label{Fgeom}}

\epsscale{0.5}
\plotone{f6.ps}
\figcaption[f6.ps]{[Ne~III]/[Ne~II] ratios (upper panel) for starburst
models with \mup=25, 30, 35, 50 and 100\msolar (from bottom to top),
and \lir/\llyc~line ratios (lower panel) for starburst models with
(from top to bottom) \mup=25, 30, 35, 50 and 100\msolar.  For each
\mup, three homogeneous-cluster models are shown as solid lines, for
timescales of \tsc= 1, 5, and 20 Myr.  For the neon ratios (upper
panel), the \tsc=1 Myr curves drop off most rapidly, and for the
\lbol/\llyc ratios (lower panel), the \tsc=1 Myr curves rise most
rapidly.  For the heterogeneous ensemble models (\S \ref{SSlf}) the
\tsc=1 and 20~Myr curves are shown as a dashed line for each value of
\mup. The vertical bar on the right side of each panel indicates the
range of measured ratios in the starburst sample, with the black
horizontal line indicating the average value for the sample. Due to
their low metallicities, NGC 5253 and IIZw40 are not included in the
average ratio value calculation.\label{Fsbmods}}

\plotone{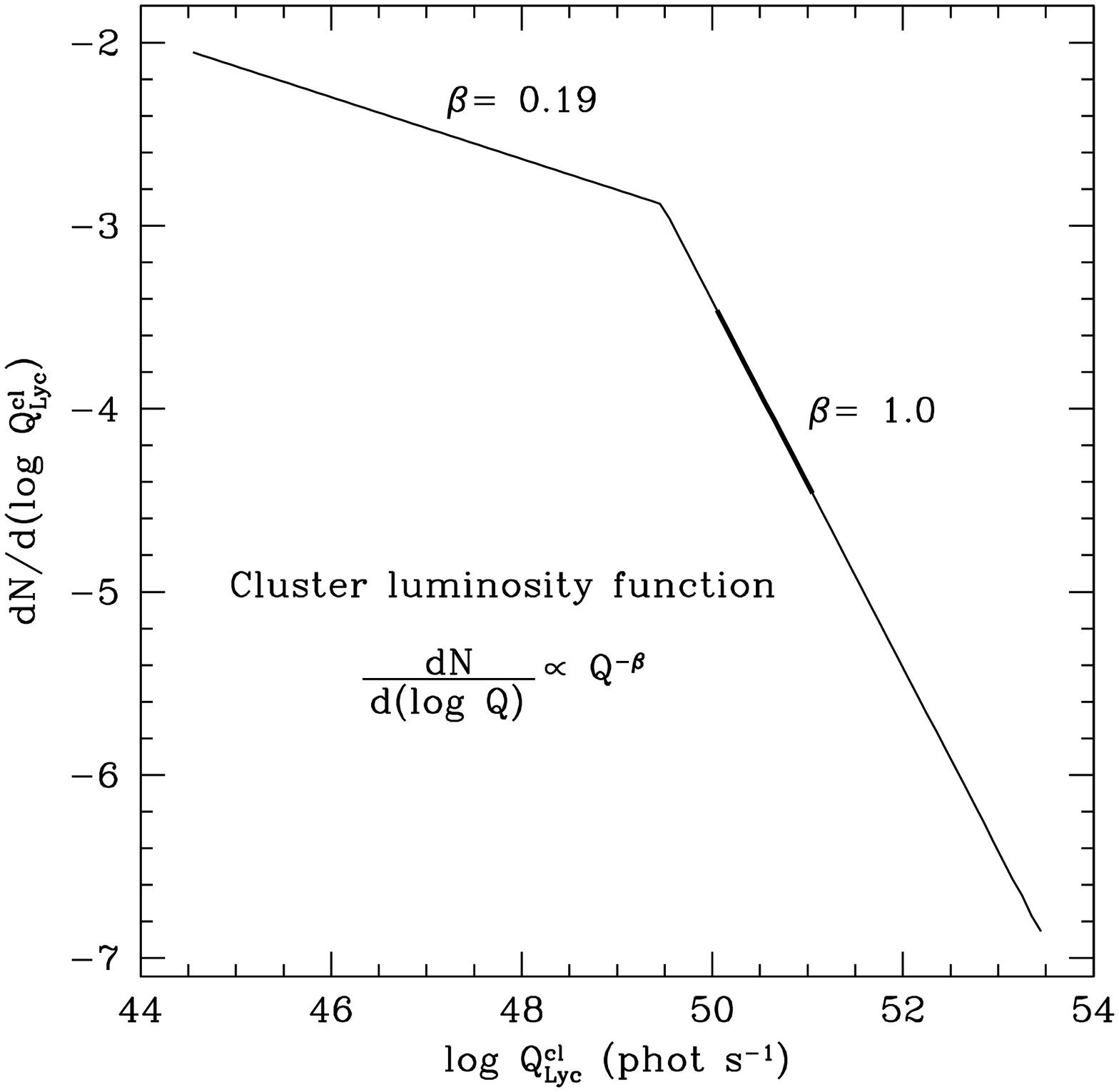}
\figcaption[f7.eps]{The form of the cluster luminosity function (in
relative units) described in \S \ref{SevolLF}, assuming $\alpha$=1.
The thick line segment indicates the range of cluster luminosity
(Q$_{Lyc}^{\rm cl}$) over which our luminosity function overlaps with
the fit to the observed luminosity function of OB associations in the
Milky Way by McKee \& Williams (1997). \label{Flf}}

\plotone{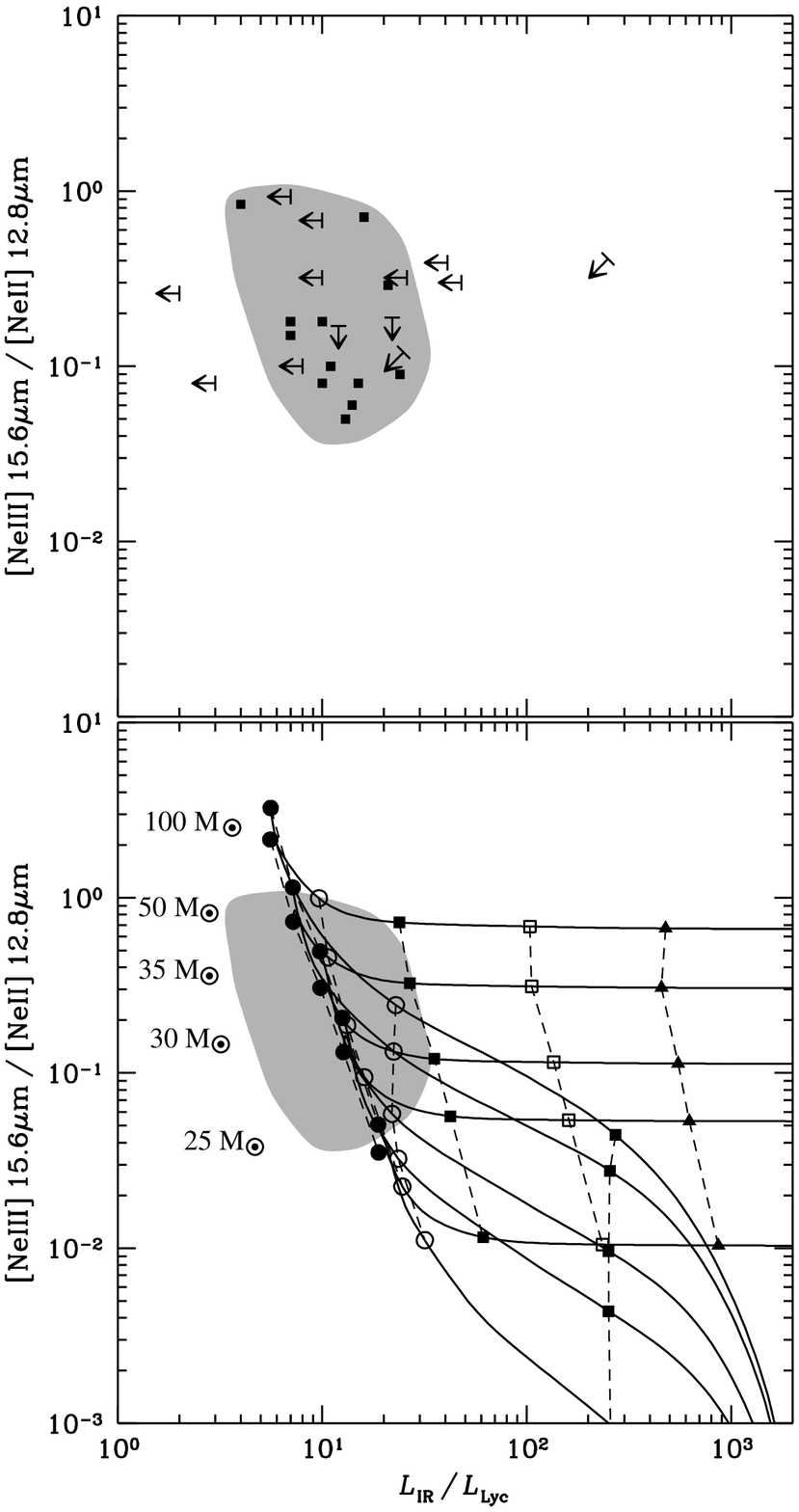}
\figcaption[f8.eps]{(Upper panel) Calculated \lir/\llyc\ ratios for
the starburst sample, plotted against their neon ratio values.  The
shaded area indicates the region of the diagram covered by all objects
for which neither ratio is an upper limit. (Lower panel) The shaded
region from the upper panel is overlaid on a set of models for
heterogeneous ensembles of clusters following the LF derived in
\S\ref{SevolLF}.  The curves are shown for the five $m_{\rm up}$ shown
in Figure \ref{Fsbmods} (\mup=25, 30, 35, 50, 100\msolar, from bottom
to top), and for burst timescales of 1~Myr and 5~Myr (lower and upper
curves, respectively, for each \mup).  The dashed lines connecting the
various dots show the ``isochrones'' in this diagram, for burst ages
of 1~Myr (filled circles), 5~Myr (open circles), 10~Myr (filled
squares), 20~Myr (open squares) and 30~Myr (filled triangles).
\label{Fsbmodsllyc}}

\plotone{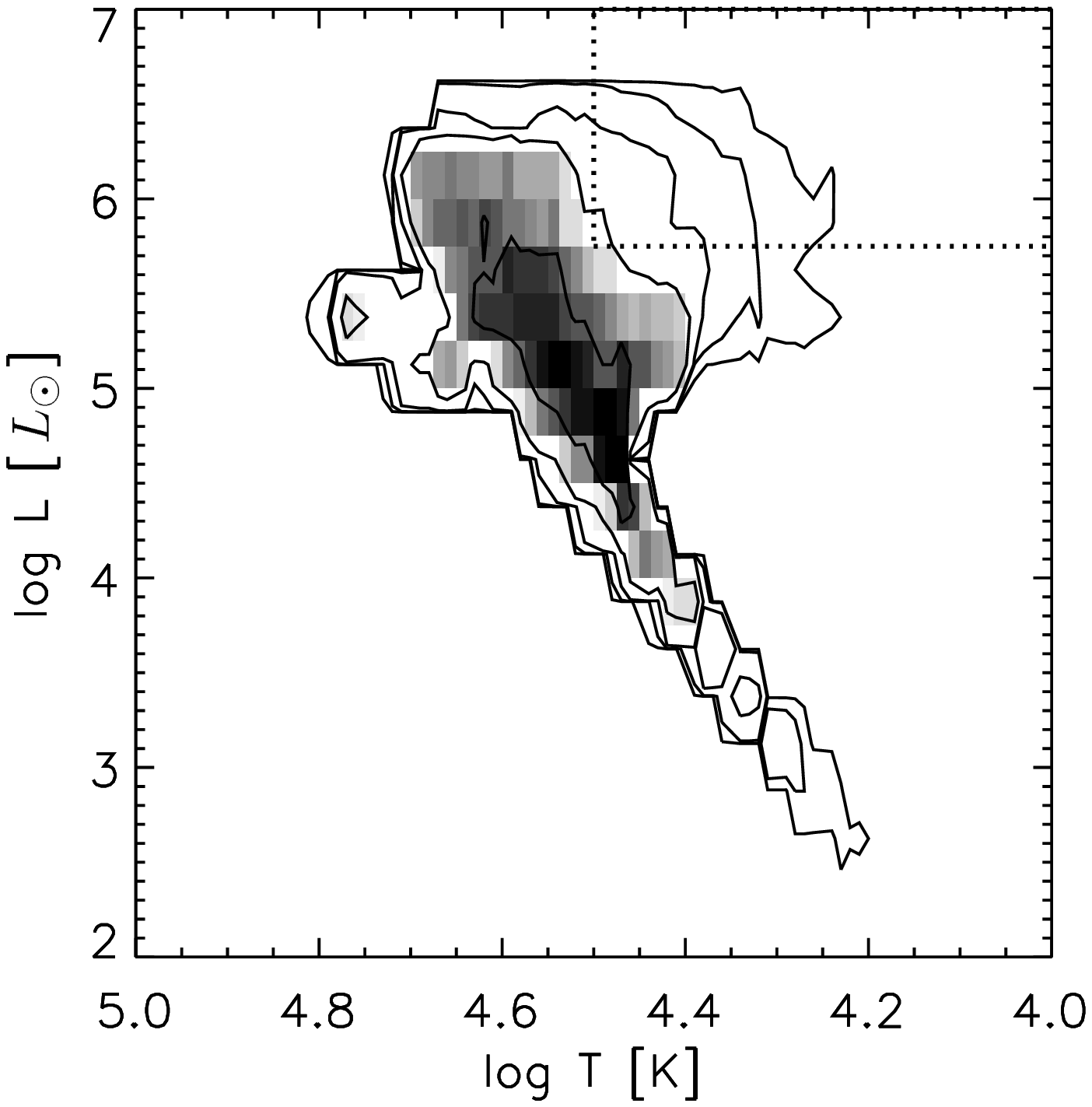}
\figcaption[f9.eps]{Contributions of different stellar types to the
total Lyman continuum luminosity of a short starburst ($t_{sc}$=1Myr)
at a burst age of 7 Myr. Contours are at 10$^{-6}$, 10$^{-5}$, ...,
10$^{-2}$ contribution of a $\Delta$ log L = 0.25, $\Delta$ log
T=0.0125 bin to the total luminosity. The greyscale coded regions are
above the 10$^{-3}$ level and contribute more than 97\% of the total
flux for this model. The box enclosed by a dotted line indicates the
region which hosts stars contributing about 50\% to the ionizing
luminosity of our Galactic Center.\label{Fgcmod}}

\plotone{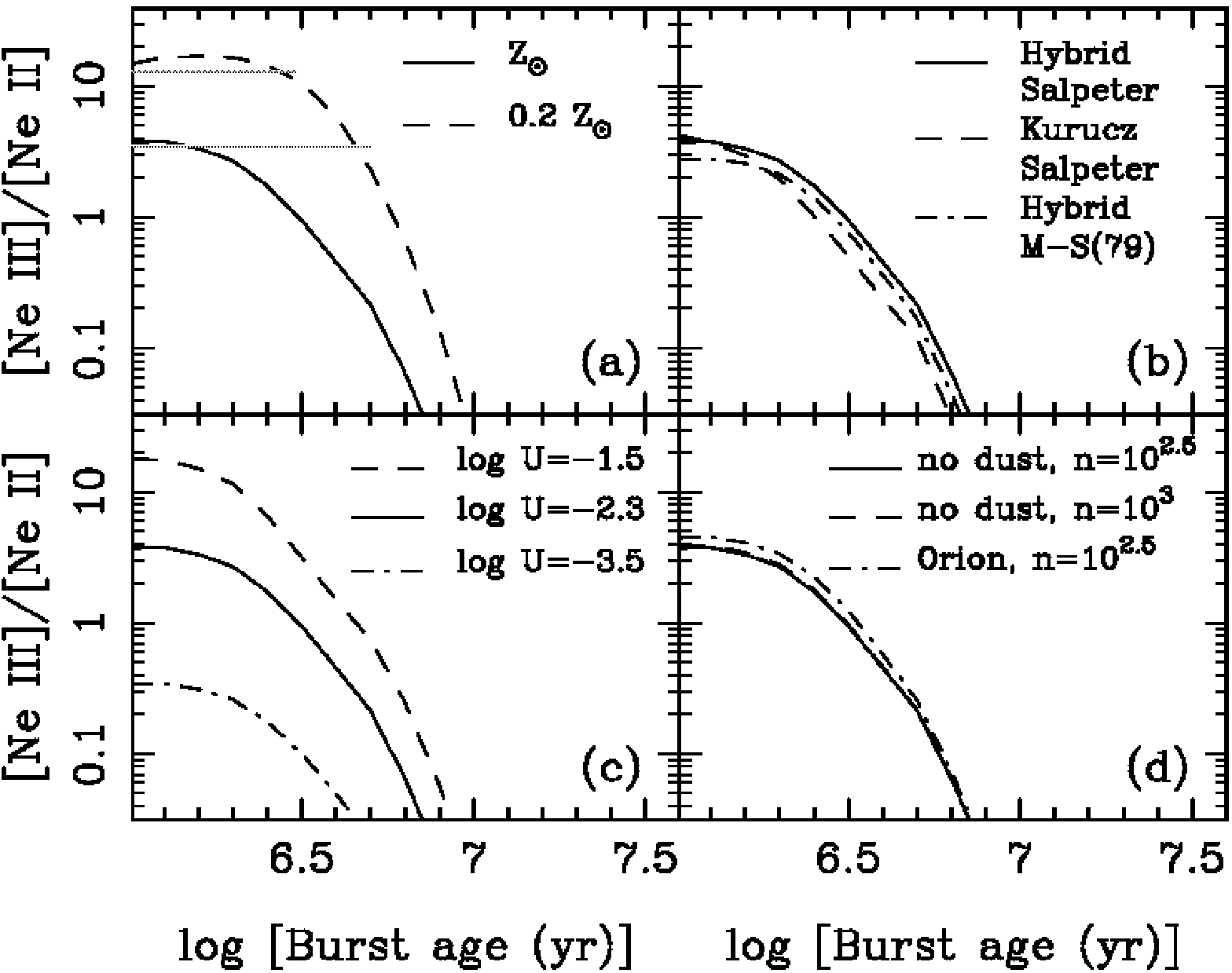}
\figcaption[f10.ps]{Variations in neon ratios due to model parameter
variations, assuming homogeneous-cluster models with \mup=100\msolar~
and \tsc=1~Myr. (a) Plot showing the neon ratios for solar
vs. subsolar (0.2~\zsolar). The measured neon ratios for IIZw40 and
NGC 5253 are indicated by the upper and lower grey lines,
respectively. (b) Neon ratios for a burst composed of SEDs from our
hybrid grid, and from a grid composed solely of Kurucz models. In
addition, the neon ratios for the hybrid grid using a Miller-Scalo IMF
are also shown.  (c) Neon ratios for log U=-1.5,-2.3, and -2.5. (d)
Neon ratios determined in environments with gas densities of 300 and
1000 cm$^{-3}$; for the case of 300 cm$^{-3}$, we also show the effect
of adding an Orion-like dust component to the CLOUDY
modeling.\label{Fvariation}}

\plotone{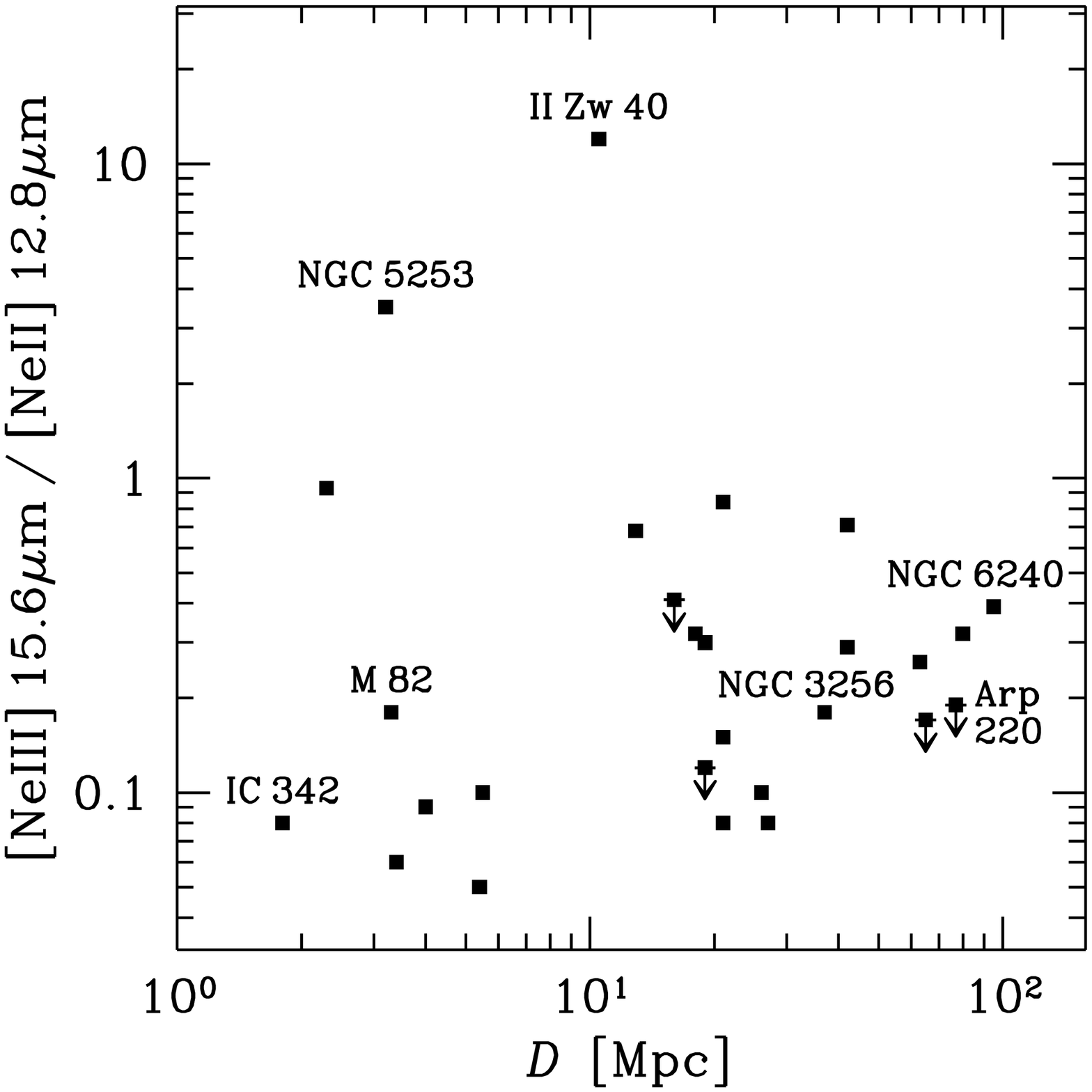}
\figcaption[f11.eps]{Plot of the
[Ne~III]/[Ne~II] ratios for the starburst sample as a function of
estimated distance.\label{Fneondist}}

\begin{deluxetable}{lcccccc}
\tablewidth{0pt}
\tablenum{1}
\tablecolumns{7}
\tableheadfrac{.05}
\tablecaption{\label{Tlines}Emission tracers from massive stars in ISO starburst sample}
\tablehead{\colhead{Galaxy} & \colhead{D }& \colhead{L$_{IR}$\tablenotemark{a}} & \colhead{F([Ne II])}  & \colhead{F([Ne III])} & \colhead{$\frac{[Ne~III]}{[Ne~II]}$} & \colhead{$\frac{\rm L_{IR}}{\rm L_{Lyc}}$\tablenotemark{j}}\\
 & \colhead{(Mpc)}& \colhead{(10$^9$~\lsolar)} & \colhead{(10$^{-19}$~W cm$^{-2}$)} & \colhead{(10$^{-19}$~W cm$^{-2}$)} & & }
\startdata
NGC 55\tablenotemark{b} & 2.3 & 0.063 & 0.53 & 0.49 &        0.93  & $\le$7\\    
NGC 253 & 3.4 & 20 & 40 & 2.5 &       0.06  & 14\\    
NGC 278\tablenotemark{b} & 12.9 & 2.5 & 0.40 & 0.24 &      0.68  & $\le$10\\    
MCG+12-02-001\tablenotemark{b} & 63 & 127 & 2.5 & 6.5 & 0.26 & $\le$2\\
IC1623A\tablenotemark{b,c} & 80\tablenotemark{h} & 312 & 1.5 & 0.46 & 0.32   & $\le$10\\
NGC 972\tablenotemark{b} & 21 & 13 & 2.1 & 0.31 & 0.15 & 7\\
NGC 986\tablenotemark{b} & 19 & $\le$27 & 1.4 & $\le$ 0.17 & $\le$ 0.12 & $\le$25\\
NGC 1084\tablenotemark{b} & 19 & $\le$30 & 0.60 & 0.18 & 0.30 & $\le$48\\
IC342 & 1.8 & 0.6 & 8.3 & 0.65 &       0.08 & 10\\     
NGC 1482\tablenotemark{b} & 26 & $\le$65 & 3.9 & 0.39 & 0.10 & $\le$8\\
NGC 1511\tablenotemark{b} & 18 & 11 & 0.51 & 0.16 & 0.32 & $\le$26\\
IRAS04296+2923\tablenotemark{b} & 27 & 17 & 2.8 & 0.23 & 0.08 & $\le$3\\
NGC 1792\tablenotemark{b} & 16 & $\le$25 & 0.24 & $\le$ 0.09 & $\le$ 0.41 & \nodata\\
IIZw40 & 10.5 & 3.1 & 0.14 & 1.7 & 12 & 3\\    
NGC 3034 (M82)  &  3.3 & 39 & 88 & 16 &  0.18  & 5 \\    
NGC 3256 & 37 & 250 & 7.6 & 1.4  &      0.18  & 7\\     
NGC 3690A\tablenotemark{d} & 42 & 408 & 3.2 & 0.93 &     0.29  & 21\\    
NGC 3690BC\tablenotemark{e} & 42 & 254 & 2.8 & 2.0 &     0.71  & 16\\    
NGC 4038/39\tablenotemark{f} & 21 & 11 & 0.77 & 0.65 &    0.84  & 4\\    
NGC 4945 & 4.0 & 15 & 8.8 & 0.75 &       0.09  & 24\\    
NGC 5236 (M83)  & 5.4 & 11 & 13.4 & 0.68 &       0.05  & 13\\    
NGC 5253 & 3.2 & 1.2 & 0.77 & 2.7 &       3.5   & 4\\    
Arp220\tablenotemark{g} & 77 & 1450 & 0.54& $<$ 0.10 &  $<$ 0.19 & 22\\  
NGC 6240 & 95 & 580 & 1.7  & 0.67 & $<$0.39\tablenotemark{h}  & $\le$41\\ 
NGC 6946\tablenotemark{b} & 5.5 & 2.3 & 2.6 & 0.27 &      0.10  & 11\\    
NGC 7469 & 65 & 290 & 2.3 & 0.4 &   $<$0.17\tablenotemark{h}  & 12\\    
NGC 7552 & 21 & 96 & 6.8 & 0.51 &      0.08  & 15\\    
\enddata
\tablenotetext{a}{Infrared ($\rm \lambda$ = 8 - 1000~\um)
luminosities computed from the point source flux densities listed in
the {\em IRAS} Faint Source Catalog, version 2.0 (1990), using the
prescription from Sanders \& Mirabel (1996).  The
infrared luminosities were further scaled to match the SWS aperture (see
\S\ref{SSobs}).  In some cases, no information was available to
provide a reliable scaling between the emission seen by IRAS and that
expected in the SWS aperture; for these cases, the total L$_{IR}$
computed from the IRAS Faint Source Catalog flux densities is
indicated as an upper limit.}
\tablenotetext{b}{From open-time sample.}
\tablenotetext{c}{Aperture center
($\alpha,\delta$)(J2000)=01$^h$07$^m$47.${\! ^s}$5, -17\arcdeg30\arcmin24.${\!}$\arcsec7}
\tablenotetext{d}{Eastern center,
($\alpha,\delta$)(J2000)=11$^h$28$^m$33.${\! ^s}$8, +58\arcdeg33\arcmin45.${\!}$\arcsec5} 
\tablenotetext{e}{Western center,
($\alpha,\delta$)(J2000)=11$^h$28$^m$31.${\! ^s}$2, +58\arcdeg33\arcmin44.${\!}$\arcsec9}
\tablenotetext{f}{Interaction region,
($\alpha,\delta$)(J2000)=12$^h$01$^m$54.${\! ^s}$9, -18\arcdeg53\arcmin02.${\!}$\arcsec5}
\tablenotetext{g}{Aperture center,
($\alpha,\delta$)(J2000)=15$^h$34$^m$57.${\! ^s}$3, 23\arcdeg30\arcmin11.${\!}$\arcsec6}
\tablenotetext{h}{NGC 6240 and NGC 7469 may have an AGN
contribution to the [Ne III] line emission; therefore, we regard the
obserbed ratios as upper limits. See \S\ref{SSother} for more information.}
\tablenotetext{j}{L$_{Lyc}$ used in this ratio is the average of the
values determined from HI recombination lines and from MIR neon
lines. See \S\ref{SSbollyc}.}
\end{deluxetable}

\addtocounter{table}{1}

\begin{table}
\begin{center}
\caption{\label{Torigpaul}Pauldrach Cornerstone models}
\begin{tabular}{cll}
\tableline
Stage & T$_{eff}$ & log g \\
 & (1000 K) & (m s$^{-2}$)\\
\tableline\tableline
Main sequence & 25 & 4.0 \\
 & 30 & 4.0 \\
 & 35 & 4.0 \\
 & 40 & 4.0 \\
 & 45 & 4.0 \\
 & 50 & 4.0 \\
 & 55 & 4.0 \\
 & 60 & 4.0 \\
Supergiant & 25 & 2.75 \\
 & 30 & 3.0 \\
 & 35 & 3.2 \\
 & 40 & 3.4 \\
 & 45 & 3.6 \\
 & 50 & 3.8 \\
\tableline
\end{tabular}
\end{center}
\end{table}

\begin{table}
\begin{center}
\caption{\label{Tmodelpars}Model parameters used in creating composite
  starburst SEDs}
\begin{tabular}{ll}
 Parameter & Value or range of values\\
\tableline\tableline
Initial mass function (IMF) & Salpeter (1955), dN/dm=m$^{-2.35}$ \\
Upper mass cutoff (\mup)& 25, 30, 35, 50, 100\msolar\\
Lower mass cutoff (\mlow)& 1\msolar\\
Burst age (t$_b$)& 1-50~Myr\\
Burst timescale (t$_{sc}$) & 1-20~Myr\\
\tableline
\end{tabular}
\end{center}
\end{table}

\end{document}